\documentclass[]{interact}

\usepackage{epstopdf}
\usepackage[caption=false]{subfig}

\usepackage{comment}

\usepackage[natbibapa,nodoi]{apacite}
	\usepackage{amssymb}
	\usepackage{soul}
	\usepackage{booktabs}
	\usepackage{tikz}
	\usepackage{enumitem}
	\usepackage{wrapfig}

\theoremstyle{plain}

\theoremstyle{definition}

\theoremstyle{remark}

\begin{document}

	\title{Informing Users: Effects of Notification Properties and User Characteristics on Sharing Attitudes}

	\author{
		\name{Yefim Shulman\textsuperscript{a,}$^\ast$\thanks{$^\ast$Corresponding author: Yefim Shulman, email: [efimshulman@mail.tau.ac.il], ORCID: 0000-0002-3163-9726}, Agnieszka Kitkowska\textsuperscript{b,}$^\dag$\thanks{$^\dag$ORCID: 0000-0001-7384-4552} and Joachim Meyer\textsuperscript{a,}$^\ddag$\thanks{$^\ddag$ORCID: 0000-0002-1801-9987}}
		\affil{\textsuperscript{a} Tel Aviv University, 6139001 Tel Aviv, Israel; \textsuperscript{b} Karlstad University, 651 88 Karlstad, Sweden}
	}
		
	\maketitle	
		
	\begin{abstract}
		Information sharing on social networks is ubiquitous, intuitive, and occasionally accidental. However, people may be unaware of the potential negative consequences of disclosures, such as reputational damages. Yet, people use social networks to disclose information about themselves or others, advised only by their own experiences and the context-invariant informed consent mechanism. In two online experiments ($N=515$ and $N=765$), we investigated how to aid informed sharing decisions and associate them with the potential outcomes via notifications. Based on the measurements of sharing attitudes, our results showed that the effectiveness of informing the users via notifications may depend on the timing, content, and layout of the notifications, as well as on the users’ curiosity and rational cognitive style, motivating information processing. Furthermore, positive emotions may result in disregard of important information. We discuss the implications for user privacy and self-presentation. We provide recommendations on privacy-supporting system design and suggest directions for further research.
	\end{abstract}
		

	\section{Introduction}
\label{sec:intro}

Many online user actions can lead to the disclosure of personal information about themselves and (or) others. Yet, people often lack information about the consequences of such actions (e.g., granting permissions to apps upon installation, browsing a website, or having a technology-mediated conversation online). Without a correct understanding of the potential effects of information sharing, users might expose themselves and others to unnecessary risks. A simple example can be a scenario when a job applicant, at some point in one's life, posts information about one's drinking habits on social networks. If a job recruiter accesses this information, it can become disqualifying for the applicant because it creates a not necessarily correctly interpreted image of substance abuse.

One form of informing users of online systems about what may happen to their information is through privacy policies (or other data protection-related documentation). However, to learn from such documents, users would have to have a certain level of expertise and enough time at hand to carefully inspect this broad and context-invariant material. The ``informed consent'' mechanism puts the duty of being ``informed'' onto the end-users, which is a suboptimal solution, disproportionally burdening them~\citep{martin2016prinotmat,good20047noticenotice,obar2020blotiitppatosp,steinfeld2016ppoitrak}. Different approaches have been taken to better inform user decision-making: from nudging user choices \citep{caraban2019waystonudge,acquisti2017nudges} to allowing users to customize their own privacy settings \citep{adjerid2019choicearchitecture,ZHANG2019personcusomaz}. However, when many interventions come short, a way to improve this and aid user decision-making is to associate the consequences with the actions, which can be done through relevant notifications, specific and concise \citep{pinder2018digbehchaninter,murmann2021effectivenotifications}. Otherwise, one leaves people to their own devices, hoping that learning from experience will result in the correct mental models, even though with the rapid development of technology, these mental models can quickly become irrelevant.

To ensure that the privacy-related information is presented to the users appropriately, scholars and policymakers defined a set of data protection goals (GDPR, \citeauthor{eu:gdpr}; CCPA, \citeauthor{ca:ccpa}). Transparency is one of the primary data protection goals~\citep{Hansen2015protgoals}, enabling users to make a better-informed decision, which is necessary to provide users with control over their personal information~\citep{shulman2019ipc}. Yet, transparency alone might not be sufficient to ensure users' control~\citep{adjerid2013limtransp}. Practical intervenability --- another data protection goal~\citep{Hansen2015protgoals} --- is crucial for users' control to be effective. However, online services mostly keep informing their users about their information collection and processing in the traditional and insufficient format of privacy policies that overwhelm users, seldom fulfilling legal requirements \citep[e.g., under the GDPR,][]{nouwens2020darkpatterns, obar2020blotiitppatosp} in the context of transparency and intervenability. Thus, to benefit both users and online service providers, a better way of informing users must be identified. Onscreen notifications can be a way to provide users with feedback about realized and potential outcomes of their actions~\citep{shulman2019ipc,Bellotti1993designinubicomp}, improving users' control over their personal information sharing.

Although privacy notifications are not a new concept, and some research has been conducted to assess their effects on users' behavior~\citep{Utz2019, Jackson2018}, scarce attention has been given to the interplay between the effects of notifications and other factors that may influence users' behavior. Findings from psychology and behavioral sciences have shown that the decision-making process may be affected by external stimuli, as well as by characteristics intrinsic to the individual~\citep{Loewenstein2003, Lerner2015a, kahneman1991anomalies, Evans2013}. Applying this to the context of information sharing while interacting with online services, users --- the receivers of information --- may process feedback (e.g., notifications) differently, depending on their stable and momentary predispositions (e.g., personality characteristics, emotional state). However, little is known about how different properties of privacy-related notifications may affect online sharing decisions in the presence of, and accounting for such predispositions.

With this paper, we aim to learn more about notifications as a means of providing feedback, informing users about the privacy consequences of their online sharing decisions. We look at individual differences affecting information processing, such as curiosity and cognitive style, and at the effects of the momentary affective states. We focus on social networks, as sharing information there is a ubiquitous action that reveals personal information. We argue that notifications are only useful when they are relevant to the user's current action. We focus on three UI design parameters regarding feedback: the way the information is presented to users (i.e., notification layout); the time, at which notifications appear on the screen (i.e., notification timing); and the message they convey (i.e., message content). 

	\section{Background}
\label{sec:background}

One way to aid users' decision-making regarding sharing their personal information builds upon the concept of nudging~\citep{thaler2004savemore,sunstein2014}, whereby altering a given choice architecture may influence people's decisions in the desired way. Research in privacy-related decision-making has considered both the benefits (e.g., protecting user privacy without restricting user autonomy) and risks (e.g., ``dark patterns'', nudging towards over-sharing) of nudging privacy decisions \citep{mirsch2017dnudging,acquisti2017nudges,nouwens2020darkpatterns}. Even though nudges helping users not to over-share information may be effective \citep[for instance,][]{wang2014nudgingfieldtrial,harbach2014personalriskcomm}, their effect may be momentary and diminish over time \citep{pinder2018digbehchaninter}. 

Another way to inform users is to provide them with privacy indicators, assigning privacy scores to services or outlining their privacy practices \citep[e.g.,][]{egelman2009timingplacement,patil2015interrupt}. This approach has its own limitations resulting in disregard of indicators due to various factors, such as habituation effect \citep{vance2018habituation}, over-confidence \citep{reeder2018expsamplstourtbwif}, and unwillingness to perform additional actions \citep{akawe2013warningland}, perhaps, due to (but not limited to) the default effect~\citep{dinner2011defaulteffect}, and the status quo bias~\citep{samuelson1988statusquo,kahneman1991anomalies}.

Users may also define their own information-sharing settings. Yet, when sharing and disclosure actions are ongoing and occur after the initial settings' configuration, users tend not to reevaluate and adjust their settings~\citep{adjerid2019choicearchitecture}, while the data protection, processing, and collection practices may have changed on the service provider's side, or users' own self-presentation motivations may have changed. Notifications may serve as an additional reminder to the users, allowing them to align the existing settings with their changing preferences. Notifications may also update the users' knowledge regarding the current online service provider's practices. 

We investigate the effects of UI design parameters (notification properties), individual differences in information processing (curiosity, cognitive style), and affective states on users' \textit{intention to give personal information} (IGPI) and \textit{privacy preferences}. Following the framework of the theories of reasoned action and planned behavior \citep{ajzen2001attitudes}, we operationalize the \textit{intention to give personal information} as a representation of a behavioral intention (an ``attitude toward a behavior'' \citep{ajzen2005attitudesonbehaviors} to share or disclose personal information, which may result in actual behavior --- sharing and disclosure. Privacy preferences should be closely related to social interactions on online social networks (posting) that inherently involve information sharing and disclosure. We focus on two privacy preferences: a self-presentation motivated one (\textit{restricting visibility of posts in online social networks}, PRPV), and a pro-social one (\textit{confirming posting with involved friends}, PCPF). Throughout the paper, we refer to these three attitudes as ``information sharing'' attitudes (or ``privacy attitudes'') for brevity. 

\subsection{Notification Layout}
\label{sec:b:PRI:display}

Before choosing how to respond to a notification, users form initial impressions, based on the notification design. User attention tends to be drawn to images and icons, while uncluttered and structured layouts promote scanning and reduce perceptual load \citep{sutcliffe2008gtmavaadwur}. Iconized messages can be preferable and cause less annoyance, compared with other visual design solutions \citep{tasse2016guawalmaw}. Better aesthetics positively affect user perceptions of content, usability, and attractiveness of web-based designs, even when their effects on human performance may remain equivocal \citep{Thielsch2019eioteowaoupidvt}. Empirical online privacy research demonstrated that users' comprehension and the ease of use of privacy information could be improved with concise messages \citep{Cranor2006usintforpriagents}. Both lengthy textual descriptions and cluttered icons in privacy policies may not draw user attention and tend to be overlooked \citep{Angulo2012tousapripodispman}. Icons should not clutter the design layout, and they should also be self-explanatory \citep{Siljee2015pritranspatterns}. 

We focus here on one aspect of the visual design of notifications, namely the way information is structured, i.e., the information layout. This allows us to test ideas posited (but not systematically addressed, to the best of our knowledge) in the literature on information layout in the design for privacy. Additionally, we want to extend the research on user information processing in interactions with privacy-related notifications. Finally, this may allow us to test the boundaries of effects arising merely from manipulating the structure of information layout. Therefore, we ask:

\medskip

\noindent \textit{RQ1:} How does the information layout in the notifications affect the intention to give personal information and privacy preferences (restricting post visibility and confirming posting with involved friends)?

\subsection{Notification Timing}
\label{sec:b:PRI:timing}

The timing at which notifications appear is viewed as an important determinant of user responses to notifications from online systems \citep{Kunzler2017efficnotif}, including their disclosure decisions \citep{schaub2015desnotic}. Privacy researchers approached the timing of notifications, or, more generally speaking, feedback, from several directions: as an interruption of a primary task, defined by the time delay and consequent amount of accumulated information \citep{patil2015interrupt}; as a factor of privacy-related information comprehension before or during the interaction with an app \citep{balebako2015timingsalience}; or as a confounded ``stage'' of a cost-benefit analysis, affecting decisions about purchases \citep{egelman2009timingplacement}. 

We investigate the effects of the timing of notifications in relation to the decision timing (delivered before or after the user makes a decision) because the timing of the notification may affect the responses to privacy-related feedback. To the best of our knowledge, relations between such timing and privacy attitudes have not been empirically studied, yet. Neither was timing appreciated enough as a tool connecting actions with outcomes to aid user decision-making. Therefore, we ask:

\medskip

\noindent \textit{RQ2:} How does the notification timing in relation to decision timing affect the intention to give personal information and privacy preferences (restricting post visibility and confirming posting with involved friends)?

\subsection{Message Content}
\label{sec:b:PRI:content}
Notifications not only provide information, but they may also disrupt some primary activity. As with privacy policies, users may skip or disregard the messages the notifications and warnings convey. For instance, warnings from browsers can be ignored for multiple reasons, from habituation \citep{vance2018habituation} to misunderstandings and (over-)confidence \citep{reeder2018expsamplstourtbwif}, and users tend not to follow links to make use of additional information \citep{akawe2013warningland}. Simultaneously, the interruption itself may affect human performance \citep{anderson2018samsuce} and user disclosure decisions \citep{adjerid2013limtransp}. 

We focus on decoupling the effect of the meaningful content of notifications from that of their interruptiveness to investigate the effects of content on sharing tendencies. The study of the effects of notification content in such a simplistic form can provide more evidence on whether and how people consider or disregard notification content. It will also show whether the notification content affects attitudes. Finally, it allows us to make sure the users do not ignore notifications altogether for any reason. This configuration of content, coupled with layout and timing, has not been studied before, to the best of our knowledge. Therefore, we ask:

\medskip

\noindent \textit{RQ3:} How does the message content of the notification affect the intention to give personal information and privacy preferences (restricting post visibility and confirming posting with involved friends)?

\subsection{Individual Differences and Information Processing}
\label{sec:b:indiv_diff}

Individual differences may affect perceptions and responses to warnings \citep{Meyer2004Conisswarn}, which may also be relevant for privacy notifications. Personality traits \citep{Junglas2008perstraitsconcern,BANSAL2016contextpersonmattertrust}, trust \citep{joinson2010ptsdo,BANSAL2016contextpersonmattertrust}, stable privacy attitudes \citep{Junglas2008perstraitsconcern,joinson2010ptsdo}, risk-taking and decision style \citep{egelman2015predprisecatt,coventry2016perssocframpdm} were studied in their relation with each other and privacy behaviors. However, little attention has been given to the role of cognitive style and curiosity (both of which characterize human information processing) in relation to privacy attitudes, even though cognitive style might affect cost-benefit considerations, as well as the intention to disclose information \citep{kehr2015thinkstylesprideci}, and curiosity has been shown to moderate the effects of perceived control on privacy comprehension \citep{kitkowska2020epttvdopnetioccaa}. 

\subsubsection{Cognitive (Thinking) Style: Cognitive-Experiential Self-Theory}
\label{sec:b:IDIP:CEST}

Cognitive-experiential self-theory (CEST) is a dual-process theory of personality \citep{Epstein2012CEST}. It posits that human information processing (i.e., cognitive style) relies on two interacting systems: a rational system (conscious, intentional, analytic, weakly related to affect) and an experiential system (preconscious, automatic, associative, strongly related to affect). Individual differences in cognitive styles may be assessed through the \textit{Need for Cognition} (NFC): rational cognitive style, based on the tendency to reason analytically about the experiential world \citep{cacioppo1982NFC}; and \textit{Faith in Intuition} (FII): experiential cognitive style, an essential part of CEST \citep{Epstein1996IndDifThinkStyles}. Individual cognitive style may affect people's judgments and decisions \citep[for instance,][]{SHILOH2002,biswas2009}, yet research on privacy-related decision-making has not dedicated much attention to the implications of CEST.

\subsubsection{Curiosity}
\label{sec:b:IDIP:curiosity}
Curiosity --- as information seeking that is not motivated by an extrinsic reward --- is a major driver of human behavior \citep{Loewenstein1994psychocurio,GOTTLIEB2013iscacnm}. In psychology, the information gap theory explains curiosity as a mismatch between what an individual knows and what they would like to learn to eliminate the feeling of deprivation of knowledge \citep[a knowledge gap,][]{Loewenstein1994psychocurio}. In decision-making, curiosity may have a stronger effect than regret aversion when the outcomes of choice options are uncertain \citep{VANDIJK2007curikilregre}. The study of curiosity in privacy contexts is important, because research on privacy nudges has focused on helping users avoid regrettable disclosures \citep{acquisti2017nudges}. However, little is known about the relations between curiosity, privacy attitudes, and information sharing.

In this paper, we study the relation between cognitive style and privacy attitudes, in line with \citet{kehr2015thinkstylesprideci}, and extend their results to the context of sharing on social media. We also extend the approach of \citet{kitkowska2020epttvdopnetioccaa}, and study the relation between curiosity and privacy attitudes. We ask:

\medskip

\noindent \textit{RQ4:} How are the individual differences in cognitive style and curiosity related to the intention to give personal information and privacy preferences (restricting post visibility and confirming posting with involved friends)?

\subsection{Affective States}
\label{sec:b:IDIP:context}

Theories proposed by psychologists and behavioral scientists, for instance, \textit{affect-as-information} or \textit{feeling-as-information}, imply, that \textit{affect} or \textit{affective states} (here, as an umbrella term for emotions) may influence people's decisions \citep{Schwarz2012,Schwarz2007,Clore2001}. Affective reactions may arise from the interaction with external stimuli, including contextual dependency or the way that information is presented to an individual. The resulting affective state may influence the decision outcome. For instance, people may perceive a situation as safer when being in a positive state and rely on their beliefs and attitudes. In contrast, in a negative state, people might rely on more complex cognitive processing. The negative state, signaling that the situation is unsafe, results in a focus on details and attention to new information that needs to be incorporated into the decision-making process. Some of the past work investigated the role of affect in privacy-related decision-making. For instance, \citet{Coopamootoo2017} found that positive emotions may spur sharing attitudes, increasing risks to privacy, while negative emotions lead to privacy-protective attitudes. Further,  \citet{kitkowska2020epttvdopnetioccaa} identified that affect resulting from interactions with privacy-related stimuli (e.g., privacy notices) might modify information disclosure. Having considered the theoretical assumptions and empirical findings from previous research, we aim to extend the literature on affect in privacy-related decision-making and hypothesize that affect might influence users' information sharing attitudes. Hence, we ask:

\medskip

\noindent \textit{RQ5:} Do intention to give personal information and privacy preferences (restricting post visibility and confirming posting with involved friends) differ, depending on the participants' affective state? 
	\section{Experiment 1}
\label{sec:experiment1}

\subsection{Method}
\label{sec:e1:method}
To investigate how individual differences and notification properties inform people and influence personal information sharing, we conducted Experiment 1. We hypothesized that the layout, timing, and message content of notifications should affect the attitudes toward sharing personal information.

The online experiment introduced a fictitious app \textit{PromotMe} that would suggest potentially popular posts, based on users' recently recorded online activities, to serve users' self-presentation on social networks. The participants then saw a fictitious suggested post with potentially sensitive content and answered several questions, measuring their information sharing attitudes. The experiment's scenario (Appendix~\ref{sec:app:exp_scenario}) asked the participants to imagine that the suggested post was based on their own photos taken during a recent travel, and that the post was to be published on their personal page on a social network website.\footnote{The experiment did not refer to any particular online social network at any point, as the social network aspect was used to provide a general background for the experimental scenario. We did not want to focus the participants' attention on any particular service, so that the participants would be able to provide responses having a social network service of their own choice in mind.} The notifications with different layouts and message content appeared at different times during the experiment. Additionally, two psychometric measurements of individual differences (Section~\ref{sec:e1:m:materials}) were taken either before or after the experiment to control for potential priming and fatigue effects that might have occurred due to answering relatively long questionnaires (20 questions in total in both scales). The experiment was designed based on the results of an exploratory study (Section~\ref{sec:e1:m:pilot}).

\subsubsection{Experimental Design}
\label{sec:e1:m:design}

\begin{figure}
    \centering
        \subfloat[Notification with \textit{Simple text} and \textit{Privacy} message]{\includegraphics[height=2.8cm]{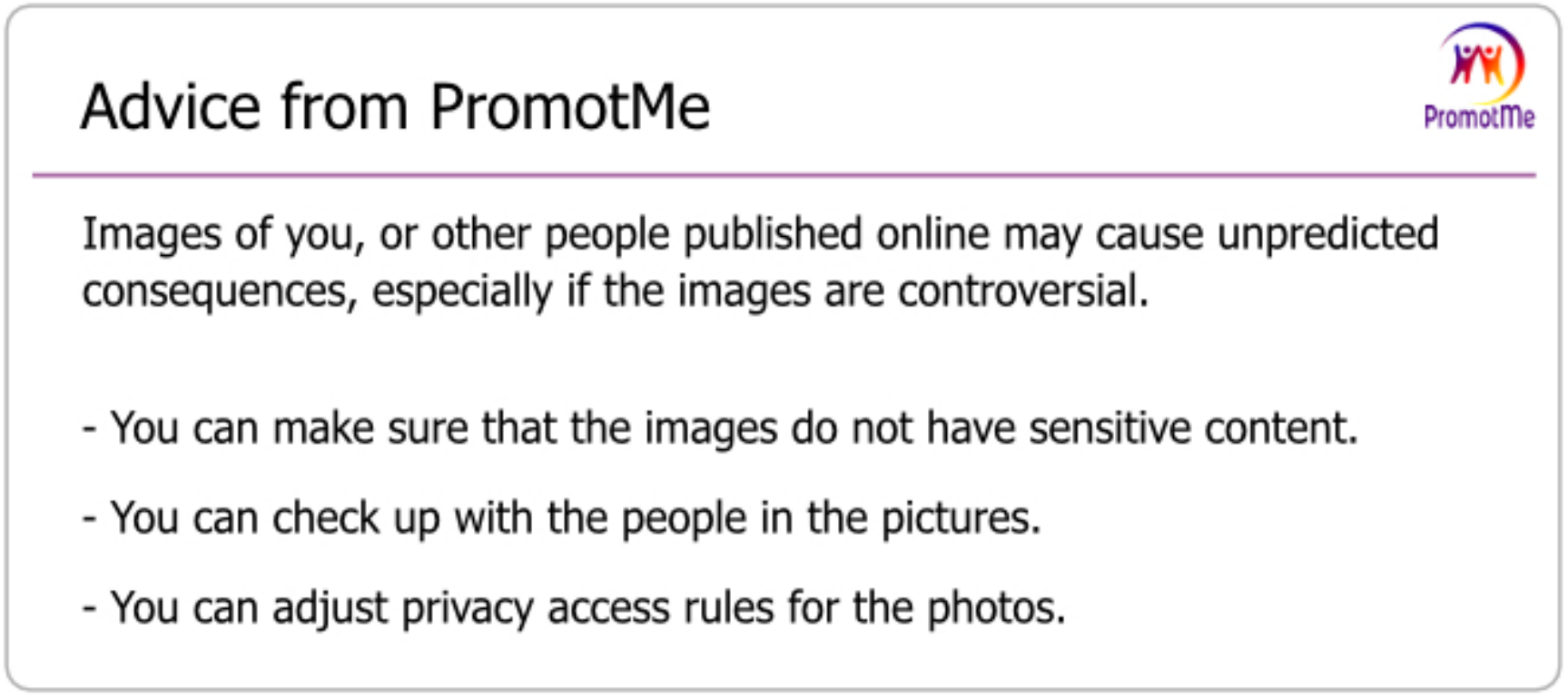}}\hspace{5pt}
        \label{fig:e1:m:text_priv}
        \subfloat[Notification with \textit{Icons} and \textit{Privacy} message]{\includegraphics[height=3.6cm]{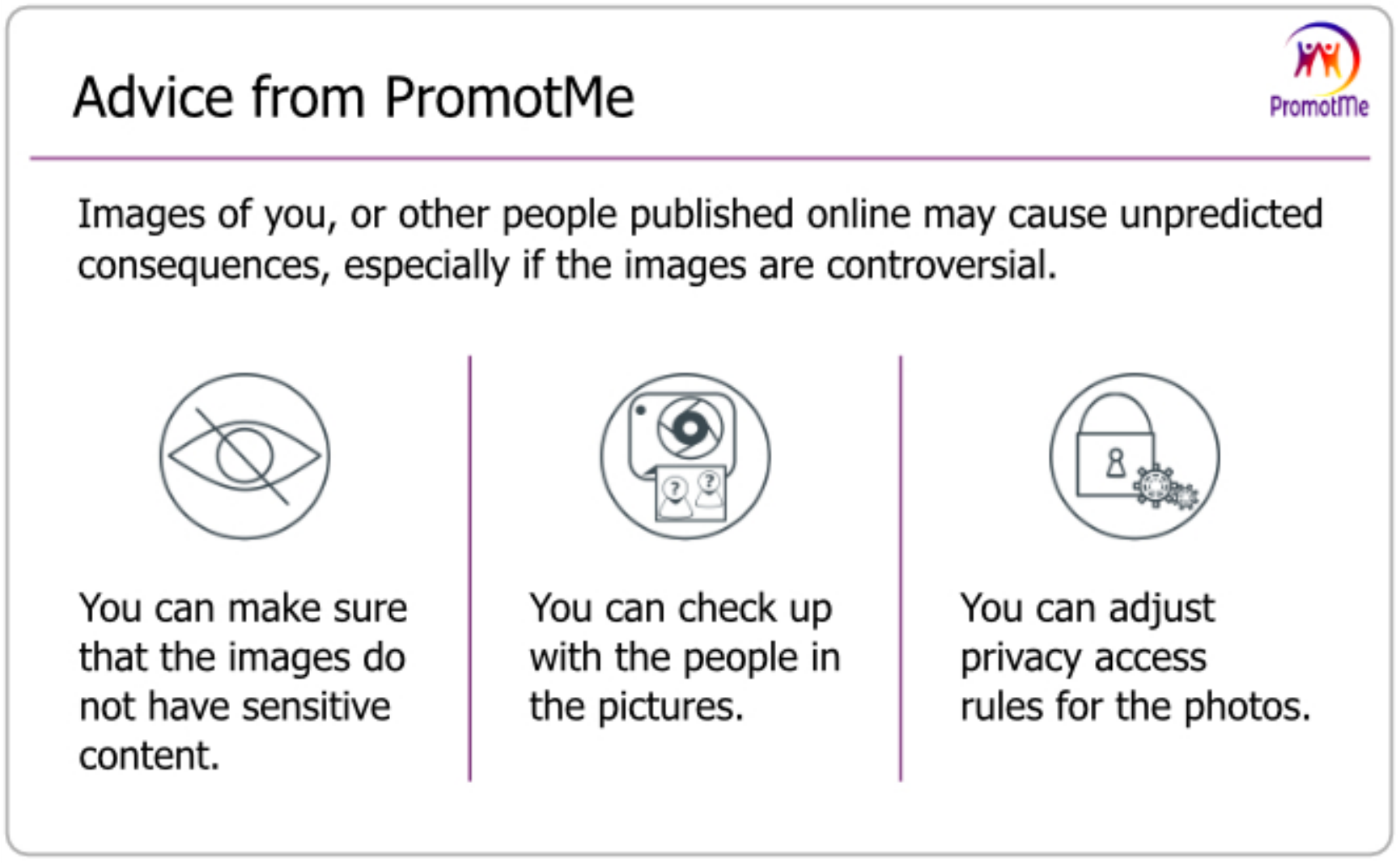}}
        \label{fig:e1:m:icons_priv}
    \caption{Examples of notification designs used in the experiment (as viewed on web-platforms)}
    \label{fig:e1:m:indications}
\end{figure}

We employed a $2\times2\times2$ full factorial experimental design, creating eight independent groups, with random assignment of participants to groups. Table~\ref{tab:e1:m:indep_vars} describes the independent variables in the experiment and introduces the shorthands used throughout the next sections of the paper to indicate the levels and variables more concisely. Fig.~\ref{fig:e1:m:indications} shows examples of notification designs (indications). Appendix~\ref{sec:app:notificitions} contains the notification designs for all combinations of the independent variables as viewed on web or mobile platforms. The randomized order of scale presentation required doubling each independent group and resulted in the twofold increase in the number of participants required for the $2\times2\times2$ experiment design.

\begin{table}[tbh]
    \centering
    \tbl{Independent variables}
   {\begin{tabular}{ll}
         \toprule
         \multicolumn{1}{c}{Variable, RQ (\textit{Shorthand})} & \multicolumn{1}{c}{Levels (\textit{Shorthands})} \\
         \midrule
            A: Notification layout, RQ1 & 1. Simple informative text \textit{(Simple text)} \\
            \hspace{3ex}(\textit{Layout}) & 2. Icons accompanied by informative text (\textit{Icons}) \\[1ex]
            B: Notification timing, RQ2 & 1. Before suggesting the post (\textit{Before}) \\
            \hspace{3ex}(\textit{Timing}) & 2. After suggesting the post (\textit{After}) \\[1ex]
            C: Message content, RQ3 & 1. Privacy implications and recommendations --- treatment group (\textit{Privacy}) \\
            \hspace{3ex}(\textit{Content}) & 2. Default settings of the fictitious app --- control group (\textit{Neutral}) \\
         \bottomrule
    \end{tabular}
    }
    \label{tab:e1:m:indep_vars}
\end{table}

We measured several constructs (dependent variables) based on participants' responses:
\begin{itemize}[noitemsep]
    \item \textit{Intention to give personal information} (IGPI). The instrument was adapted from \citet{malhotra2004iuipc} with minor adjustments. We employed four items measured on a 7-point semantic differential scale.
    \item \textit{Preference to restrict the post's visibility} (PRPV), using one ad hoc item, offering seven response options, increasing in the restrictiveness of the privacy settings.
    \item \textit{Preference to confirm posting with involved friends} (PCPF), using one ad hoc item, offering five response options, increasing in the extent of approval (social accountability), needed from friends whose personal information the post contained.
    \item \textit{Affective state}. In an open-ended question, we asked participants to share any feelings they  might have had during the experiment. 
\end{itemize}

More details regarding the measurements of the dependent variables can be found in Appendix~\ref{sec:app:DVs}.

We also controlled for the effects of several other constructs (covariates): the individual differences --- curiosity and cognitive style (Section~\ref{sec:e1:m:materials} for description and Appendix~\ref{sec:app:ind_diff} for details); and the experiences --- the number of recalled notification types, and the number of recalled online privacy violations (Appendix~\ref{sec:app:experiences}).

\subsubsection{Procedure}
\label{sec:e1:m:procedure}

\begin{figure}
    \centering
    \includegraphics[width=0.98\textwidth]{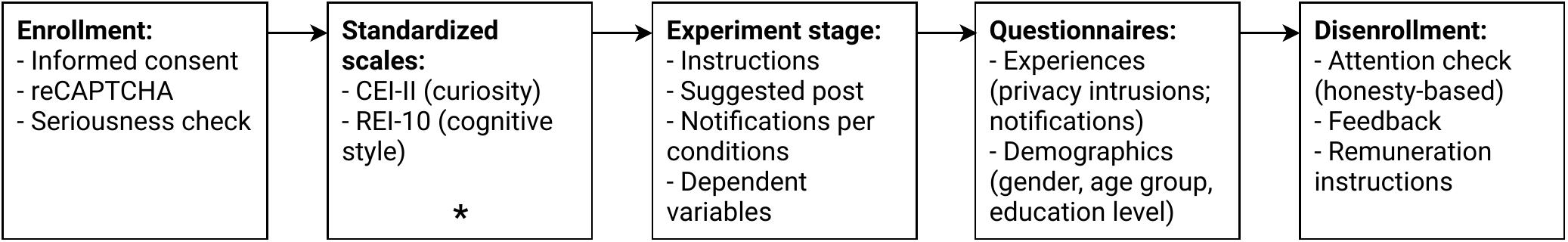}
    \caption{The experiment flow. *Standardized scales were shown either before or after the Experiment stage.}\label{fig:e1:m:exp_flow}
\end{figure}

The experiment contained five stages, as shown in Fig.~\ref{fig:e1:m:exp_flow}. 

\begin{itemize}[noitemsep, leftmargin=.0 cm]
    \item[] \textit{Enrollment:} After acknowledging the informed consent form, the participants were asked to solve a reCAPTCHA\footnote{A version of CAPTCHA by Google LLC.}, and a simple mathematical equation as a seriousness check and a safeguard against automated software agents.
    \item[] \textit{Standardized scales:} Half of the participants were asked to answer the CEI-II and, then, the REI-10 scale before the Experiment. The other half was asked to answer these inventories after the Experiment, but before the Questionnaires.
    \item[] \textit{Experiment:} This stage included the instruction screen, the post suggested by a fictitious app \textit{PromotMe}, the screens containing notifications (Fig.~\ref{fig:e1:m:indications}) per randomly assigned condition, and the measures of the dependent variables.
    \item[] \textit{Questionnaires:} The participants answered questions about their experiences with privacy intrusions, warnings and notifications from online systems, and some demographic information after the Experiment stage or after the Standardized scales stage.
    \item[] \textit{Disenrollment:} The final screen of the experiment contained an honesty-based attention check and an optional feedback form. Upon finishing the experiment, the participants were redirected to collect their remuneration.

\end{itemize}

\subsubsection{Materials}
\label{sec:e1:m:materials}

The experiment was built, using Qualtrics\footnote{Qualtrics LLC, Provo, Utah, US.}. Participants accessed the experiment online via a link, using their own devices. The experiment contained the notifications, suggested post, two scales measuring individual differences, a set of questions measuring the dependent variables (Section~\ref{sec:e1:m:design}, Appendix~\ref{sec:app:DVs}), and a questionnaire of experiences (Appendix~\ref{sec:app:experiences}) and demographics.

\textit{Notifications and suggested post.} We developed two sets of notifications (four for web-based participation and four adapted for mobile screens (Appendix~\ref{sec:app:notificitions}), and two compositions of the suggested post (Fig.~\ref{fig:suggested_post} in Appendix~\ref{sec:app:suggested_post}) to make sure the experiment's user interface design was responsive. Table~\ref{tab:e1:m:inditexts} presents the exact phrasings of the messages in the notifications.

\begin{table}
    \centering
    \tbl{Messages conveyed through notifications}
  {\begin{tabular}{lll}
         \toprule
         \multicolumn{1}{c}{Message section} & \multicolumn{2}{c}{Message \textit{Content} of notifications} \\
         \cmidrule{2-3}
         & \multicolumn{1}{c}{\textit{Privacy} (treatment group)} & \multicolumn{1}{c}{\textit{Neutral} (control group)} \\
         \midrule
         Header & ``Advice from PromotMe'' & ``Advice from PromotMe'' \\
         \midrule
         General statement & ``Images of you, or other people published & ``The PromotMe app will be creating \\
         &  \hspace{3ex}online may cause unpredicted conse- &  \hspace{3ex}and suggesting you posts based \\
         & \hspace{3ex}quences, especially if the images are  &  \hspace{3ex}on the following settings.'' \\
         & \hspace{3ex}controversial.'' &\\
         \midrule
         Action statements & ``You can make sure that the images & ``Screen language: English (United \\
         &  \hspace{3ex}do not have sensitive content.'' & \hspace{3ex}States).'' \\
         & ``You can check up with the people & ``Interface layout: Default.'' \\
         & \hspace{3ex}in the pictures.'' & \\
         & ``You can adjust privacy access rules & ``Screen resolution: User's Custom.'' \\
         & \hspace{3ex}for the photos.'' & \\
         \bottomrule
    \end{tabular}
    }
    \label{tab:e1:m:inditexts}
\end{table}

\textit{Scales measuring individual differences.} We employed two validated scales that assess individual differences:
\begin{itemize}[noitemsep]
    \item We used the \textit{Curiosity and Exploration Inventory} (CEI-II from \citet{Kashdan2009curiosity}, Appendix~\ref{sec:app:ind_diff:CEIII}) to measure curiosity. It consists of two dimensions: \textit{stretching} as the motivation to seek new information and experiences (5 items); and \textit{embracing} as the ``willingness to embrace the novel, uncertain, and unpredictable nature of everyday life'' \citep[][p.995 --- 5 items]{Kashdan2009curiosity}. 
    \item We used the \textit{Rational-Experiential Inventory} (REI-10 or ``REI-short'' from \citet{Epstein1996REI10}, Appendix~\ref{sec:app:ind_diff:REI10}) to assess cognitive style. REI-10 assesses two cognitive styles: the conscious, rational cognitive style (5 items, based on the \textit{Need for Cognition} measure); and the pre-conscious, experiential style (5 items, based on the \textit{Faith in Intuition} measure).
\end{itemize}

\subsubsection{Participants}
\label{sec:e1:m:participants}

We used CloudResearch panels to recruit participants for the experiment. The CloudResearch users were English speakers and at least 18 years of age. We estimated the required sample size using G*Power~3 \citep{Faul2007gpower}. Given our experiment design parameters and accounting for potential errors (due to, for instance, seriousness or attention check failures), and after closely inspecting the data, we retained $N=515$ complete responses. All subjects were rewarded for their participation according to the CloudResearch rates\footnote{CloudResearch panels drew users from different platforms, which were not revealed to us. CloudResearch policy states, ``Upon completion of the study, you will receive compensation in the amount that you have agreed to with the platform through which you entered this survey.''}. Involvement in our research was voluntary and could be terminated by the participants at any stage of the experiment, with no negative consequences. The informed consent form containing this information (among other instructions) and the experiment itself were approved through the university's ethical review process. We did not collect any sensitive personal information (i.e., special categories of personal data under the GDPR \citep[][Article 9]{eu:gdpr}. Table~\ref{tab:e1e2:m:participants} presents the sample demographics.

\begin{table}[ht]
    \centering
    \tbl{Sample demographics in the two experiments}
    {\begin{tabular}{l l r r r r}
         \toprule
         \multicolumn{1}{c}{Demographic} & \multicolumn{1}{c}{Level} & \multicolumn{2}{c}{Experiment 1} & \multicolumn{2}{c}{Experiment 2} \\ 
         \cmidrule{3-6}
         && \multicolumn{1}{c}{\textbf{\textit{n}}} & \multicolumn{1}{c}{\textbf{$\%$}} & \multicolumn{1}{c}{\textbf{\textit{n}}} & \multicolumn{1}{c}{\textbf{$\%$}} \\
         \midrule
         Gender & Female & 311 & 60.4 & 464 & 60.7 \\
         & Male & 202 & 39.2 & 291 & 38.0 \\
         & Non-binary / Other & 1 & 0.2 & 8 & 1.0 \\
         & Preferred not to say & 1 & 0.2 & 2 & 0.3 \\
            \midrule
         Age cohort & 18--24 & 47 & 9.1 & 88 & 11.5 \\
         & 25--34 & 63 & 12.2 & 133 & 17.4 \\
         & 35--44 & 58 & 11.3 & 128 & 16.7 \\
         & 45--54 & 61 & 11.8 & 72 & 9.4 \\ 
         & 55--64 & 110 & 21.4 & 89 & 11.7 \\
         & 65 or older & 176 & 34.2 & 251 & 32.8 \\
         & Preferred not to say & 0 & 0.0 & 4 & 0.5 \\
           \midrule
         Highest com- & No formal schooling / education & 0 & 0.0 & 6 & 0.8 \\
         \hspace{3ex}pleted level & Some high school, no diploma & 12 & 2.3 & 26 & 3.4 \\
         \hspace{3ex}of education & High school diploma or an equivalent & 76 & 14.8 & 200 & 26.1 \\
         & Some college credit, no degree & 109 & 21.2 & 156 & 20.4 \\
         & Trade, technical, vocational training & 64 & 12.4 & 21 & 2.7 \\
         & Associate's degree or an equivalent & 3 & 0.6 & 76 & 9.9 \\
         & Bachelor's degree or an equivalent & 168 & 32.6 & 161 & 21.1 \\
         & Master's degree or an equivalent & 67 & 13.0 & 81 & 10.6 \\
         & Doctorate degree or an equivalent & 16 & 3.1 & 29 & 3.8 \\
         & Other & 0 & 0.0 & 2 & 0.3 \\
         & Preferred not to say & 0 & 0.0 & 7 & 0.9 \\
           \midrule
         Total && 515 && 765 & \\
         \bottomrule
    \end{tabular}
    }
    \label{tab:e1e2:m:participants}
\end{table}

We presented participants with a list of 13 online privacy violations and with a list of 15 types of warnings and notifications (Appendix~\ref{sec:app:experiences}), and asked whether they could remember encountering any of those. Regarding online privacy violations, $36.3\%$ ($n=187$) of participants recalled at least one item from the list ($M=1.51$, $SD=1.61$, range: $0-11$, $N=515$). Regarding familiarity with warnings and notifications, $9.3\%$ ($n=48$) of participants recalled encountering ``in the last month'' at least one notification from our list ($M=5.44$, $SD=3.69$, range: $0-15$, $N=512$). As  $n=3$ participants provided a non-response to the question regarding familiarity with warnings, further inferential analyses in Section~\ref{sec:e1:r:main_results} are conducted on $N=512$ full responses. Overall, the distribution of participants across the between-subjects groups in the experiment was balanced: range of $30-33$ participants per group (sixteen groups, if the order of scale presentation is considered as part of experiment design --- Section~\ref{sec:e1:m:design} for clarifications); or $62-66$ participants per group (eight groups, if the order of scale presentation is not considered). The between-subjects groups did not differ in the demographics significantly.

\subsection{Exploratory Study}
\label{sec:e1:m:pilot}

We pre-tested the manipulation and the notification designs in an exploratory study to inform the current research design.  We collected $N=52$ responses from undergraduate students. The results provided evidence for the effects of such notification parameters as \textit{Layout} and \textit{Timing}. Message \textit{Content} was conceptualized and implemented differently (as text intelligibility implemented with fonts and colors), did not have the \textit{Neutral} level and did not seem to be effective. Based on the overall results, we adjusted the dependent variables (switched the ad hoc IGPI measure to the validated one and opted for the single- instead of multi-item measures for the privacy preferences) and modified message \textit{Content} both conceptually (as a research question), and practically (as having two levels --- treatment vs. control groups --- instead of having several permutations based on differences in fonts and colors). 
	\subsection{Results}
\label{sec:e1:results}

Before conducting the main data analysis, we tested the reliability and validity of the psychometric scales. Next, we examined the correlations to ensure that the appropriate tests are applied for the main data analysis.

\subsubsection{Measurement Instruments}
\label{sec:e1:r:meas_ind_diff}

We measured the IGPI with a validated 4-item scale, and the individual differences with CEI-II for curiosity and REI-10 for cognitive style, each containing ten items.

\textit{Intention to give personal information.} We tested the IGPI with a principal component analysis (PCA). It showed that all four items loaded a single factor, as expected, explaining $73.10\%$ of cumulative variance in the items. Sampling adequacy was acceptable, according to a Kaiser-Meyer-Olkin (KMO) overall measure of~$.76$ \citep{Kaiser1974IFS}. Bartlett's test of sphericity was significant at $<.001$, so the items were suitable for exploratory factor analysis (EFA). The reliability of measurement was acceptable, estimated with Cronbach's $\alpha=.87$ (higher than the recommended level of .70 \citep{Gliem2003CARLTS}. Removal of any of the four items would not increase the reliability. Therefore, the IGPI score was calculated as an average of all four items.

\textit{Curiosity.} We used CEI-II to measure two dimensions of curiosity: \textit{stretching} and \textit{embracing}. We used PCA, and the items initially loaded two factors with eigenvalues of $4.59$ and $3.85$ after rotation. The KMO measure was $.93$, and Bartlett's test of sphericity was significant at $p<.001$, meaning that both the sample and items were suitable for EFA. All items' extracted communalities were higher than .61, meaning that items were related to each other and there was no evidence for hidden factors \citep{costello2005bestefa}. However, four items cross-loaded both factors with medium to high loadings, and seven items had medium to high correlations with both factors. The exclusion of the four cross-loading items resulted in an even weaker factor structure and decreased the reliability of the remaining six measures. Therefore, we decided to use all ten items to calculate a single curiosity score, as the two dimensions described curiosity as a single trait. Cronbach's $\alpha=.91$ for the ten items measure indicated strong construct reliability (excluding any item would not improve the reliability).

\textit{Cognitive style.} We assessed the participants' cognitive style along two dimensions: \textit{rational style} and \textit{experiential style}, measured with REI-10. The KMO measure was $.78$, and Bartlett's test of sphericity was significant at $<.001$. The initial PCA produced three factors. The \textit{experiential style} loaded its five items correctly, while the \textit{rational style} split into two factors. Further reliability analyses and PCAs revealed that one item from the \textit{rational style} dimension reduces the reliability of its sub-scale. Its exclusion from the analysis (Appendix~\ref{sec:app:ind_diff:REI10}) resulted in 9 items splitting between two dimensions, as intended, explaining $34.25\%$ and $26.55\%$ of the total variance in the items with a clean pattern/component matrix. The final calculation of the score for the \textit{experiential style} measure included all five FII items with a reliability of Cronbach's $\alpha=.84$. The final calculation of the score for the \textit{rational style} measure included four of the NFC items with a reliability of Cronbach's $\alpha=.77$. All items contributed to the reliability of both sub-scales.

\subsubsection{Descriptive Analysis}
\label{sec:e1:r:descriptives}

We report select descriptive statistics to show relations between variables in the sample (Table~\ref{tab:e1:r:corrs}). We found a moderate negative association between the IGPI and the PRPV and PCPF (both were moderately positively associated). This indicates that if people were inclined to share information in the first place, the privacy preferences were less strict, and vice versa --- if people were prone to privacy considerations, the intention to share decreased.

Whether CEI-II and REI-10 were shown before or after the main experiment was not significantly associated with any variables of interest, meaning that the order of scale presentation may be excluded from further analyses, but we verify that in Section~\ref{sec:e1:r:main_results}.

\begin{table}
    \centering
    \tbl{Pearson correlations: intention to give personal information (IGPI), preference to restrict the post's visibility (PRPV), preference to confirm posting with involved friends (PCPF), curiosity (CSE), REI-10's need for cognition (NFC), REI-10's faith in intuition (FII), number of recalled privacy violations (NPV), number of recalled warnings and notifications (NWN).}
    {\begin{tabular}{l l l l l l l l}
         \toprule
         & PRPV & PCPF & CSE & NFC & FII & NPV & NWN \\
         \midrule
         IGPI & $-.56$*** & $-.42$*** & .33*** & .03 & .06 & .10* & $-.11$* \\
         PRPV & 1 & .54*** & $-.17$*** & .07 & $-.05$ & $-.11$* & .07 \\
         PCPF && 1 & $-.10$* & .05 & $-.06$ & $-.07$ & .06 \\
         CSE &&& 1 & .45*** & .22*** & .15*** & .03 \\
         NFC &&&& 1 & .15** & .08 & .13** \\
         FII &&&&& 1 & .01 & .03 \\
         NPV &&&&&& 1 & .35*** \\
         \bottomrule
    \end{tabular}
    }
    \tabnote{***$p<.001$, **$p<.01$ and *$p<.05$. $N=515$, except for NWN, where $N=512$.}
   \label{tab:e1:r:corrs}
\end{table}

\subsubsection{Effects of Notifications and Individual Differences}
\label{sec:e1:r:main_results}

To address the RQs (Section~\ref{sec:background}), we used an analysis of covariance (ANCOVA) on the dependent variables (IGPI, PRPV, and PCPF).\footnote{When the ANCOVA assumptions are met, it may be equivalent to multiple regression analyses \citep{tabachnick2019stats,cohen2013regres}, but allows to estimate the mean differences, simple effects and run contrasts and post hoc analysis more straightforwardly, which makes ANCOVA advisable to analyze the experiments \citep{huitema2011ANCOVA}.} Before conducting the statistical analyses, we checked the ANCOVA assumptions (linearity, homogeneity of variances, homoscedasticity, normality, and multicollinearity), which were met. To better understand the relations of the individual differences (RQ4) and other covariates with the dependent variables, we also conducted simultaneous multiple regression analyses, making sure the data met the regression assumptions (Table~\ref{tab:e1:r:regressions}).

\medskip
\textit{Intention to give personal information.} We investigate the effects the \textit{Layout}, \textit{Timing}, and \textit{Content} may have on the IGPI (first part of RQ1--RQ3), accounting for the influence of the individual differences (first part of RQ4).

We performed an ANCOVA with the IGPI as a dependent variable. Neither the order of scale presentations, nor the answer to the seriousness check question (numerical vs. verbal answer) had a significant effect on responses. The platform type (mobile vs. desktop) used by participants did not significantly affect the responses.  Therefore, the final model included several predictors: the \textit{Layout}, \textit{Timing}, and \textit{Content} as independent variables; and \textit{Curiosity}, \textit{Rational style}, \textit{Experiential style}, the number of recalled notification types, and the number of recalled online privacy violations as covariates (adjusted $R^2=.16$).\footnote{The adjusted $R^2$ estimates the linear components of the amount of variance explained by the independent variables affecting the dependent variable in the models \citep{miles2005}. It is generally interpreted as a share or a percentage.}

We found between-subjects main effects\footnote{We provide a broader note on the effect sizes in Section~\ref{sec:gd:NoteOnEffectSizes}.} of the notification \textit{Timing} (RQ2), $F(1,499)=5.23$, $p<.05$, $\eta_p^2=.010$, and \textit{Content} (RQ3), $F(1,499)=5.18$, $p<.05$, $\eta_p^2=.010$, on the participants' \textit{intention to give personal information}. The participants exposed to the notification \textit{Before} sharing expressed lower IGPI compared with those exposed to the notification \textit{After} sharing ($M=2.30$, $SE=0.09$, $95\%$ CI$[2.12-2.48]$ vs. $M=2.60$, $SE=0.09$, $95\%$ CI$[2.42-2.79]$, respectively). \textit{Privacy}-related notifications reduced the participants' IGPI compared with \textit{Neutral} notifications of the control group ($M=2.30$, $SE=0.09$, $95\%$ CI$[2.12-2.48]$ vs. $M=2.60$, $SE=0.09$, $95\%$ CI$[2.42-2.78]$, respectively). Thus, both the \textit{Privacy}-related notification and notification shown \textit{Before} sharing lead participants to be less eager to share. As to the RQ1, we found a null effect of \textit{Layout}, which indicates that in the experiment's context, the structuredness of information display might have been of lesser relevance. 

Additionally, we found significant effects of covariates. \textit{Curiosity} had a medium-to-large effect on the IGPI, $F(1,499)=64.25$, $p<.001$, $\eta_p^2=.114$ (RQ4). \textit{Rational style} had a small effect on the IGPI $F(1,499)=8.43$, $p<.01$, $\eta_p^2=.017$, while \textit{Experiential style} did not have a significant effect (RQ4). The number of notification types recalled by the participants also helped explain some variance in the IGPI, $F(1,499)=11.18$, $p<.01$, $\eta_p^2=.022$, as did the number of recalled online privacy violations, $F(1,499)=6.19$, $p<.05$, $\eta_p^2=.012$. Table~\ref{tab:e1:r:regressions} shows the directions and relative strength of the associations between the covariates and the information sharing attitudes.

\medskip
\textit{Preference to restrict the post's visibility}. Overall, participants tended to opt for more restrictive privacy settings in terms of visibility of the suggested post for different social groups (PRPV, $M=5.63$, $SD=1.62$, $Mdn=6.00$, range: $1-7$, --- the higher, the fewer users would potentially be able to see the suggested post if/when shared). We used an ANCOVA to study the effects of the \textit{Layout}, \textit{Timing}, \textit{Content}, \textit{Curiosity}, \textit{Rational cognitive style}, \textit{Experiential cognitive style}, the number of recalled notification types, and the number of recalled online privacy violations on the PRPV (adj. $R^2=.07$).

We found a between-subjects main effect of the \textit{Content} (RQ3), $F(1,499)=4.04$, $p<.05$, $\eta_p^2=.008$, on the \textit{preference to restrict the post's visibility}. The \textit{Privacy}-related notification message increased the participants' PRPV compared with the \textit{Neutral} one of the control group ($M=5.77$, $SE=0.10$, $95\%$ CI$[5.57-5.96]$ vs. $M=5.49$, $SE=0.10$, $95\%$ CI$[5.29-5.68]$, respectively), indicating that the participants were more defensive in response ot the \textit{Privacy}-related notification. As to RQ1 and RQ2, we found null effects of \textit{Layout} and \textit{Timing}, which indicates that the structuredness of information display and \textit{Timing} of notifications in the experiment may have been of lesser relevance to the PRPV.

In terms of the individual differences (RQ4), we observed significant effects of two covariates --- \textit{Curiosity}, $F(1,499)=20.59$, $p<.001$, $\eta_p^2=.040$, and \textit{Rational cognitive style}, $F(1,499)=13.05$, $p<.001$, $\eta_p^2=.025$, --- and no significant effect of \textit{Experiential style}. The number of recalled notifications types and the number of online privacy violations were both significant in the analysis of the PRPV: $F(1,499)=5.33$, $p<.05$, $\eta_p^2=.011$, and $F(1,499)=7.09$, $p<.01$, $\eta_p^2=.014$, respectively. Table~\ref{tab:e1:r:regressions} presents further detail on the covariates' relations to the information sharing attitudes.

\medskip
\textit{Preference to confirm posting with involved friends.} We built an ANCOVA model, treating the responses to the PCPF item as an interval-scaled variable. The model included the same factors and covariates as the one described previously. Our analyses did not reveal statistically or practically significant effects of the independent variables on the PCPF (RQ1--RQ3 in the part pertinent to PCPF, adj. $R^2=.02$). However, regarding the individual differences (RQ4), the data revealed significant effects of \textit{Curiosity}, $F(1,499)=6.10$, $p<.05$, $\eta_p^2=.012$ and \textit{Rational cognitive style}, $F(1,499)=4.63$, $p<.05$, $\eta_p^2=.009$, with the direction of the relations shown in Table~\ref{tab:e1:r:regressions}. Thus, the individual differences in information processing were associated with the information sharing attitudes. Interestingly, \textit{Curiosity} and \textit{Rational style} appeared to have opposite relations with the sharing attitudes consistently, while \textit{Experiential style} had no relation with any of the three sharing attitudes.

\begin{table}
    \tbl{Joint influence of the curiosity (CSE), REI-10's need for cognition (NFS), REI-10's faith in intuition (FII), number of recalled privacy violations (NPV), number of recalled warnings and notifications (NWN) on the intention to give personal information (IGPI) and the privacy preferences (PRPV and PCPF).}
    {
    \begin{tabular}{lrrrrrrrrr}
         \toprule
         Predictor & \multicolumn{3}{c}{IGPI,} & \multicolumn{3}{c}{PRPV,} & \multicolumn{3}{c}{PCPF,} \\
         & \multicolumn{3}{c}{$F(5,506)=18.25$,} & \multicolumn{3}{c}{$F(5,506)=7.86$,} & \multicolumn{3}{c}{$F(5,506)=3.22$,} \\
         & \multicolumn{3}{c}{$p<.001$, adj. $R^2=.14$} & \multicolumn{3}{c}{$p<.001$, adj. $R^2=.06$} &  \multicolumn{3}{c}{$p<.01$, adj. $R^2=.02$} \\
         \cmidrule(){2-10}
         & \multicolumn{1}{c}{\textbf{$\beta$}} & \multicolumn{1}{c}{\textbf{$t(506)$}} & \multicolumn{1}{c}{\textbf{$r_p$}} & \multicolumn{1}{c}{\textbf{$\beta$}} & \multicolumn{1}{c}{\textbf{$t(506)$}} & \multicolumn{1}{c}{\textbf{$r_p$}} & \multicolumn{1}{c}{\textbf{$\beta$}} & \multicolumn{1}{c}{\textbf{$t(506)$}} & \multicolumn{1}{c}{\textbf{$r_p$}} \\
         \midrule
         \textbf{CSE} & \textbf{.39***} & \textbf{8.19} & \textbf{.34} & \textbf{$-$.22***} & \textbf{$-$4.56} & \textbf{$-$.20} & \textbf{$-$.12*} & \textbf{$-$2.45} & \textbf{$-$.11} \\
         \textbf{NFC} & \textbf{$-$.13**} & \textbf{$-$2.90} & \textbf{$-$.13} & \textbf{.18***} & \textbf{3.64} & \textbf{.16} & \textbf{.11*} & \textbf{2.16} & \textbf{.10} \\
         FII & .00 & 0.07 & .00 & $-.03$ & $-0.73$ & $-.03$ & $-.05$ & $-1.12$ & $-.05$ \\
         \textbf{NPV} & \textbf{.10*} & \textbf{2.34} & \textbf{.10} & \textbf{$-$.12*} & \textbf{$-$2.55} & \textbf{$-$.11} & $-.09$ & $-1.81$ & $-.08$ \\
         \textbf{NWN} & \textbf{$-$.14**} & \textbf{$-$3.22} & \textbf{$-$.14} & \textbf{.10*} & \textbf{2.16} & \textbf{.10} & .08 & 1.73 & .08 \\
         \bottomrule
    \end{tabular}
    }
    \tabnote{***$p<.001$, **$p<.01$, and *$p<.05$.}
   \label{tab:e1:r:regressions}
\end{table}

\subsubsection{Affective States}
\label{sec:e1:r:qualitative}

In addition to the quantitative analysis, we conducted a qualitative analysis of the responses to the open-ended question that asked participants to describe how they felt about the experimental stimuli and the posting overall. 

First, we used WordStat to identify the most frequently used words and phrases in open-ended responses. This quantitative approach enabled better understanding of the context of the affective states that the participants were describing. The most frequently used words by our respondents were: \textit{photos} ($n=160$, $33\%$), \textit{post} ($n=140$, $29\%$) and \textit{posting} ($n=62$, $13\%$), \textit{people} ($n=56$, $11\%$), and \textit{social} \textit{media} (accounted for separately as words with $n=59$ ($12\%$) and $n=48$ ($10\%$), respectively). Most frequently used phrases in the open-ended questions were ``Social media'' ($n=51$, $10\%$) and ``post photos'' ($n=21$, $4\%$), followed by ``people in the photos'', ``promoteme notice'', ``post on social media'', ``poor taste'' accounting for approximately $1\%$ each. Both notifications and privacy issues in them were mentioned less frequently, suggesting that people were more concerned about the posts' content that might be negatively perceived, impacting self-presentation or the image of others. These quantitative results informed the design of the second experiment (Section~\ref{sec:experiment2}), where we decided to use additional suggested posts with more neutral pictures.

Next, two researchers independently analyzed the data, discussed any inconsistencies, and identified main categories within the data. We divided the 483 valid responses into categories based on affective states: \textit{positive ($n=66$), negative ($n=297$), neutral ($n=57$), multiple ($n=29$)}, and \textit{unclear ($n=34$)}. The \textit{multiple} category contained both positive and negative opinions, and the \textit{unclear} category was assigned when researchers were unable to recognize the affective notion.

To further analyze the data (RQ5), following the circumplex model of affect (posi-tive -- neutral -- negative valence), we used only \textit{positive},  \textit{neutral}, and \textit{negative} categories. Thus, we reduced the total sample size to $N=420$. We applied a mixed-method analysis and utilized categorical results as independent variables to identify a relationship between the affective states and the information sharing attitudes (IGPI, PRPV, and PCPF). We used an analysis of variance (ANOVA) or Welch ANOVA when the homogeneity of variances was violated.  Because of the unequal number of respondents in the groups, we applied bootstrapping. 

\begin{figure}
     \centering
     \subfloat[Intention to give personal information.]{\includegraphics[width=0.45\textwidth]{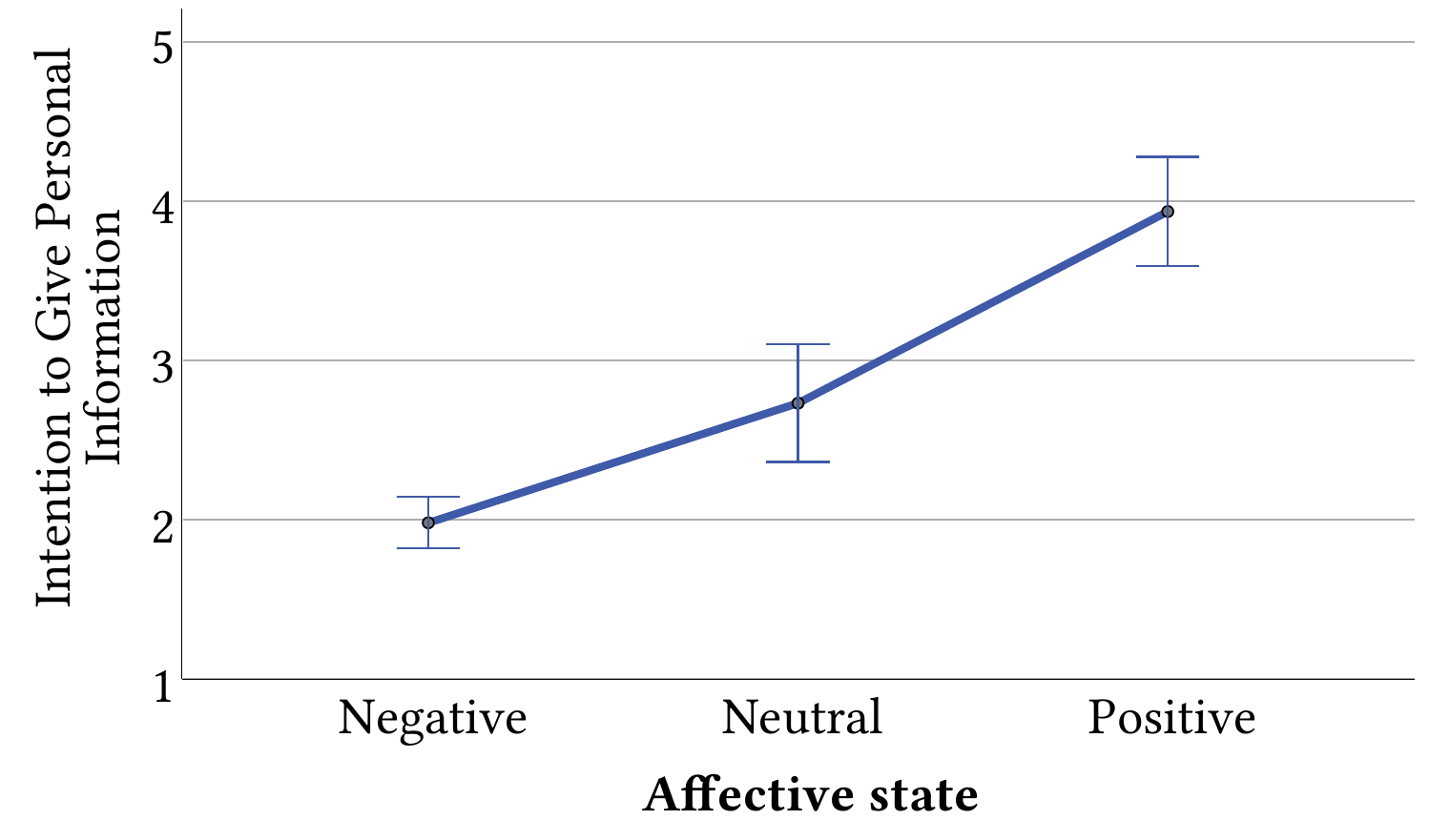}
         \label{fig:e1:r:igpi}}
     \hspace{5pt}
     \subfloat[Preference to restrict the post's visibility.]{\includegraphics[width=0.45\textwidth]{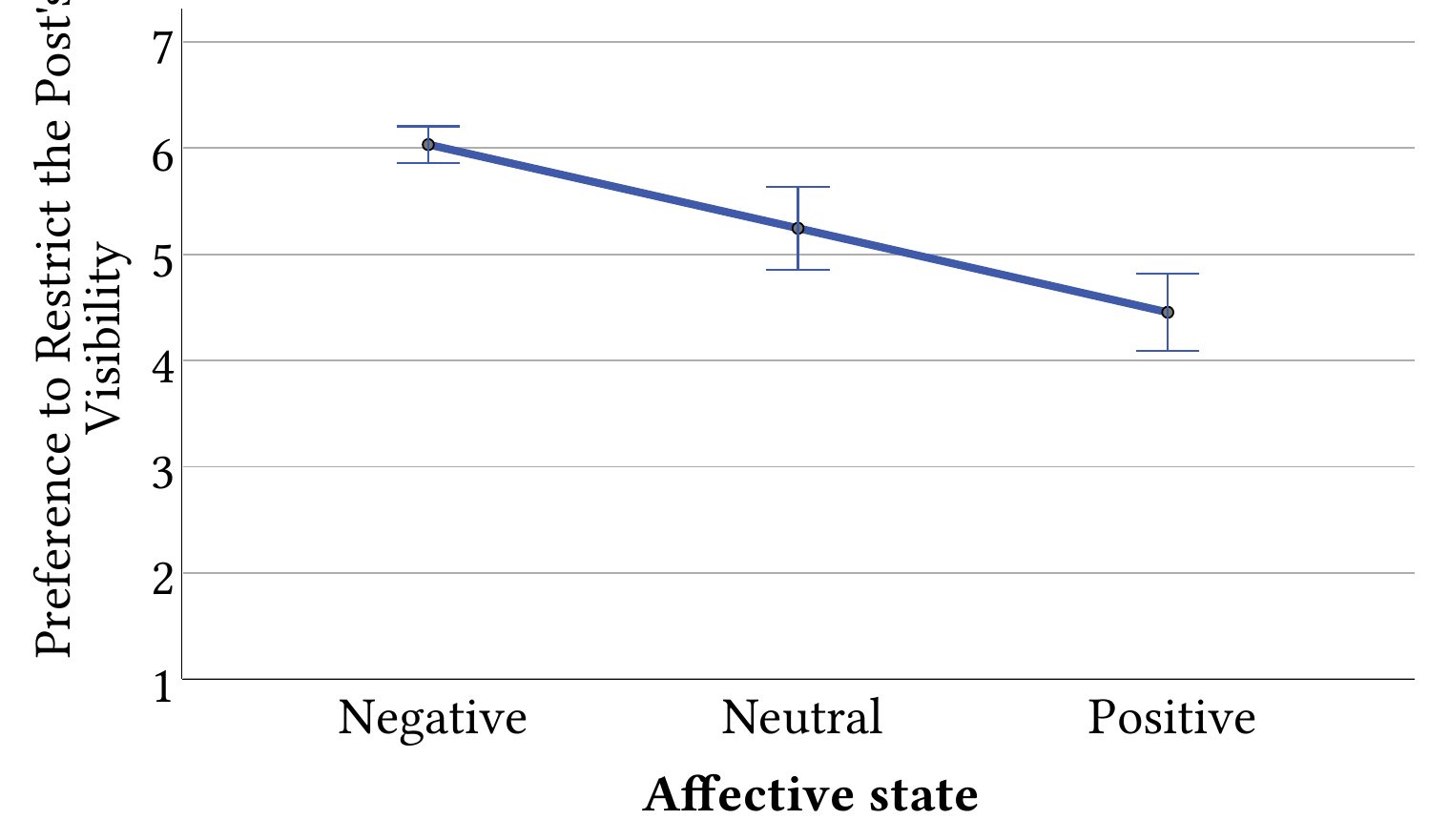}
         \label{fig:e1:r:PRPV}}
     \hfill
     \subfloat[Preference to confirm posting with involved friends.]{\includegraphics[width=0.45\textwidth]{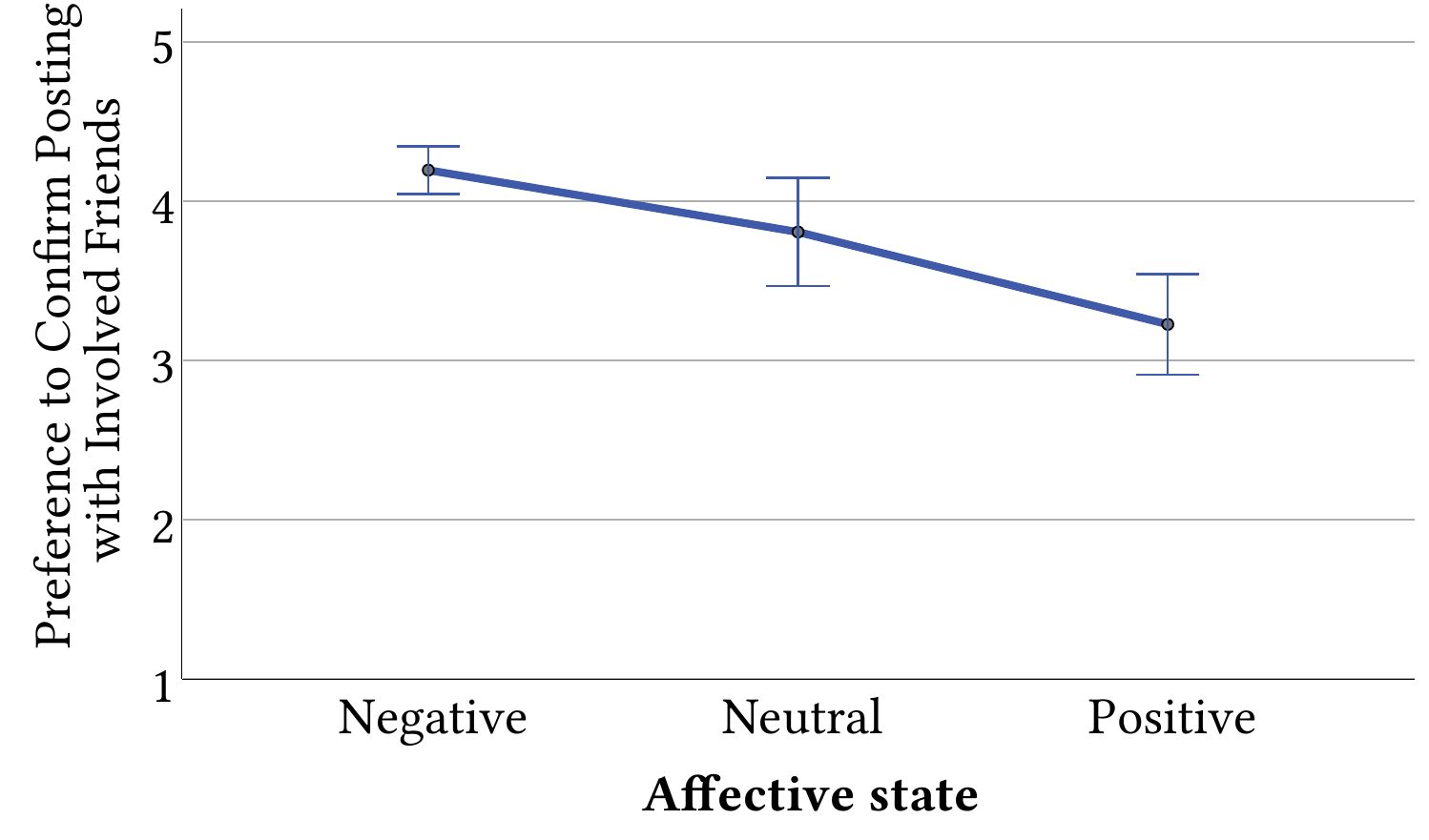}
         \label{fig:e1:r:PCPF}}

    \caption{The differences in the means of the dependent variables based on affective states. Error bars represent $95\%$ CI.}
    \label{fig:e1:r:affect}
\end{figure}

The test results showed significant between-groups differences for the \textit{intention to give personal information}, \textit{Welch's} $F(2,95.07)=38.02$, $p<.001$, \textit{est.} $\omega^2=.15$ (Figure~\ref{fig:e1:r:igpi}). We used Games-Howell bootstrapped confidence intervals to identify which groups differed significantly. Our results demonstrated significant differences between all groups: \textit{positive} vs. \textit{negative} CI$[1.5,2.4]$, \textit{positive} vs. \textit{neutral} CI$[0.6,1.8]$, \textit{neutral} vs. \textit{negative} CI$[0.2,1.2]$. The participants in the \textit{positive} state ($M=3.93$) tended to have higher IGPI than the participants in the \textit{negative} ($M=1.98$) or \textit{neutral} ($M=2.73$) states. Moreover, the participants in the \textit{negative} state seemed to have significantly lower IGPI than the people in the \textit{neutral} state.

A similar analysis testing how affect was related to the \textit{preference to restrict the post's visibility} revealed significant between-groups differences, \textit{Welch's} $F(2,102.42)=28.11, p<.001$, \textit{est.} $\omega^2=.11$ (Figure~\ref{fig:e1:r:PRPV}) between participants in \textit{positive} vs. \textit{negative} CI$[-2,-1.2]$, \textit{positive} vs. \textit{neutral} CI$[-1.4,-0.2]$, and \textit{neutral} vs. \textit{negative} CI$[-1.2,-0.3]$ affective states. Participants in a \textit{positive} affective state ($M=4.45$) had significantly lower PRPV than those in a \textit{negative} ($M=6.03$) or \textit{neutral} ($M=5.25$) states. Similarly, participants in a \textit{neutral} state had a significantly lower PRPV than those in a \textit{negative} state. 

Finally, we used the ANOVA to establish how affective state was related to the \textit{preference to confirm posting with involved friends}. We applied a standard one-way bootstrap test because the homogeneity assumption was not violated (Levene's test $p>.05$). We found significant differences between-groups, $F(2,417)=15.48, p<.001, \eta^2=.07$ (Figure~\ref{fig:e1:r:PCPF}). The Tukey post-hoc test revealed that the significant differences were between the participants in \textit{positive} vs. \textit{negative} CI$[-1.31,-0.56]$, and \textit{positive} vs. \textit{neutral} CI$[-1.1, -0.03]$ affective states. The difference between participants in \textit{negative} and \textit{neutral} states was not significant. These results imply that participants in \textit{negative} ($M=4.20$) and \textit{neutral} ($M=3.81$) affect are more likely to seek an approval from their friends before posting, than participants in a \textit{positive} ($M=3.23$) affective state.

\subsection{Discussion}
\label{sec:e1:discussion}

We investigated five RQs in an online experiment, obtaining a variety of results. 

\subsubsection{Notification Properties}
\label{sec:e1:d:notif_propert}

\smallskip
\textit{Layout.} We aimed to broaden the literature on privacy designs and find ways to ease information comprehension, as prior literature highlighted such enhancement potential \citep{waddell2016paraphraseddesign,tabassum2018incrusattcomicpolicy}. \citet{Angulo2012tousapripodispman} recommended uncluttered icons to aid user attention and information comprehension. In the exploratory study (Section~\ref{sec:e1:m:pilot}), we saw the effect of \textit{Icons} vs. \textit{Simple text}. Yet, we did not find a difference between the notifications containing \textit{Icons}, compared with \textit{Simple text} notifications in the more diverse sample of the online experiment. Perhaps this illustrates a case when the amount of information, or the situation context caused participants to pay equal attention to both experimental layouts.

\smallskip
\textit{Timing.} The \textit{Timing} (RQ2) of notifications may connect actions to outcomes. We found that it affected the IGPI, suggesting that the feedback information was acknowledged. Yet, the privacy preferences remained unaffected, so users' preferences might be relatively stable or habitual regarding with whom to share (PRPV) and whose involvement to consider (PCPF).

These results imply that the timing of feedback can make information more salient and (or) relevant for the users, aiding informed choices. We argue that such timing can help prevent regrettable actions that may have negative consequences for users' privacy and self-presentation. We advise system designers and privacy practitioners to inform users regarding the implications of their actions in a concise manner \textit{Before} they are faced with choice options.

\smallskip
\textit{Content.} We tested the effects of context-relevant \textit{Content} of notifications on the IGPI, PRPV, and PCPF and compared it to the effect of merely interruptive notifications communicating context-irrelevant content (RQ3). \textit{Content} indeed had a robust effect on the IGPI and PRPV, enhancing privacy-protective attitudes. 

Importantly, it suggests that the content of feedback can inform users and prevent regrettable actions in terms of privacy and (or) self-presentation consequences. The effect of \textit{Content} but not of $[Content \times Timing]$ does not support the previous findings in \citet{adjerid2013limtransp}, where an interruption completely negated the effect of privacy policy presentation. Perhaps the difference can be explained by the amount of information provided in policies vs. shorter notifications, duration of the decision-making stage, or simply differences between contexts of notifications and sharing in our experiment, and policies and disclosures in the experiment in \citet{adjerid2013limtransp}.

\subsubsection{Individual Differences in Information Processing}
\label{sec:e1:d:ind_diff}

In the experiment, we also addressed the potential relation between the individual differences, such as \textit{Curiosity} and \textit{Cognitive style}, and the users' IGPI and privacy preferences (RQ4).

We found that the level of curiosity during the interaction is related to the behavioral intention to share information (IGPI) and privacy preferences regarding who should be able to view the shared information (PRPV), and whose privacy should be considered (PCPF). These findings add to the existing literature, which showed that curiosity can influence information comprehension, and in moderation with perceived control may also influence the users' affective states \citep{kitkowska2020epttvdopnetioccaa}. We argue that curiosity may be related to alertness and the tendency to process information, helping better inform the users or motivating users to inform themselves.

The data also revealed that the rational cognitive style --- the Need for Cognition --- was significantly related to the IGPI, PRPV, and PCPF, but the effect size was smaller, compared with \textit{Curiosity}. Contrary to \citet{kehr2015thinkstylesprideci}, there was no significant effect of the experiential cognitive style (Faith in Intuition) on the IGPI or privacy preferences. The differences may be due to the context: we explored the sharing of potentially damaging information on social networks, while in \citet{kehr2015thinkstylesprideci}, the authors focused on a driving style support application. Perhaps car driving is indeed a more automatic process relying largely on experience, while sharing (even impulsively) usually requires time and focused effort to compose and publish a message. Thus, it seems that the rational part of cognition might be more relevant in privacy-related interactions when sharing online.

\subsubsection{Affect}
\label{sec:e1:d:affect}

The participants' affective state was related to the IGPI and both privacy preferences (RQ5). Theories, such as \textit{affect-as-information} and \textit{feeling-as-information} \citep{Schwarz2012,Schwarz2007,Clore2001}, imply that positive affect may result in lesser attention to external information. Similarly, participants in the more positive state intended to give more information, and their preferences to restrict post visibility or to confirm posting with involved friends were lower. These findings suggest that people in positive affective states may be more vulnerable to information disclosure manipulations. Thus, they might be more likely to over-disclose and deal with the consequences of regrettable disclosures. The emotional state may reduce the effects of design factors providing external information, such as notification layout and message content. We advise researchers to direct more attention to the effects of emotions in privacy decision-making.

\bigskip
Even though the free-form feedback and responses to the open-ended question regarding affect provided confidence in the effectiveness of the manipulation (notifications), participants mostly attended to the suggested post itself. The fact that we only showed a single post may limit the understanding of the generalizability of the results. We address this issue in the second experiment by introducing two other suggested posts and implementing technical and minor methodological improvements. We also look more closely at the effects of the affective states.

	\section{Experiment 2}
\label{sec:experiment2}

The second experiment expanded the scope of the first one by introducing two additional suggested posts and including validated quantitative measures of affect, taken twice instead of the single open-ended question. We also substituted REI-10 with a half of REI-40, made minor adjustments to the experiment flow, and improved some phrasing and response options in the demographic questionnaire.

\subsection{Method}
\label{sec:e2:method}

The second experiment was based on the first experiment (Section~\ref{sec:experiment1}). It introduced the same fictitious app \textit{PromotMe}. The scenario instructions and the notifications were identical to those in the first experiment. However, this time the participants saw one of three fictitious suggested posts: we kept the original suggested post from Experiment 1 and added two other variants, depicting different experiences (Appendix~\ref{sec:app:suggested_post}, Section~\ref{sec:e2:m:design}). The two psychometric measurements of individual differences were taken after the experiment, because the order of presentation of the psychometric scales did not have any effect in the first experiment. The CEI-II scale remained unchanged. However, REI-10 was substituted with a part of REI-40 (Sections~\ref{sec:e2:m:design} and \ref{sec:e2:m:materials}, and Appendices~\ref{sec:app:ind_diff:CEIII} and \ref{sec:app:ind_diff:REI40}) to be more specific regarding \textit{Rationality} and \textit{Experientiality}. REI-40 measures 4 constructs equally: \textit{Rational ability}, \textit{Rational engagement}, \textit{Experiential ability}, and \textit{Experiential engagement}, whereas REI-10 is asymmetric: it measures \textit{Rational ability} and \textit{Experiential engagement}. To shorten the questionnaire, we included only a half of REI-40, measuring only \textit{Rational} and \textit{Experiential} \textit{abilities}, to keep the symmetry. We considered the use of only a half of REI-40 sufficient, because the pairs of measures of \textit{Rationality} (i.e., \textit{ability} and \textit{engagement}) and \textit{Experientiality} (i.e., \textit{ability} and \textit{engagement}) were strongly correlated, both in the original paper and in multiple other uses of the full REI-40 \citep[for instance,][]{akinci2013AssIndDiffExpRatCogStyles,McLaughlin2014RatExpDMStudents,SLADEK2010AgGenDifRatExpCog,phillips2015ThinkStylesMetaAnalysis}. We chose the \textit{abilities}, rather than \textit{engagements}, because the items seemed to be phrased more clearly.

\subsubsection{Experimental Design}
\label{sec:e2:m:design}

The major difference, compared with Experiment 1, was the use of three suggested post variants as another between-subjects variable. Therefore, Experiment 2 had a $[2\times2\times2]\times3$ full factorial experimental design, totaling 24 independent groups with random assignment of participants to groups (the 8 groups in square brackets signify the properties of notifications taken unchanged from the first experiment). The \textit{Suggested post variants} included the original post from the first experiment (with images of partying and heavy \textit{Drinking}), the new post with images showing \textit{Gambling}, and the new post with images containing sights of a popular touristic destination --- \textit{Traveling}. 

We measured the same dependent variables as in the first experiment, apart from the \textit{Affective state}, which we this time measured using the PANAS (Positive and Negative Affect Schedule \citep{watson1988PANAS} twice over the course of the study to capture the possible change of the affective states (Section~\ref{sec:e2:m:procedure} and \ref{sec:e2:m:materials}, and Appendix~\ref{sec:app:DVs:affect:2}). Using the standardized scale also allowed us to measure affect without the researchers' interpretations of participants' verbal responses.

The covariates were the same as in the first experiment, apart from the aforementioned substitution of REI-10 by the \textit{Ability} half of REI-40.

\subsubsection{Procedure}
\label{sec:e2:m:procedure}

\begin{figure}
    \centering
    \includegraphics[width=0.98\textwidth]{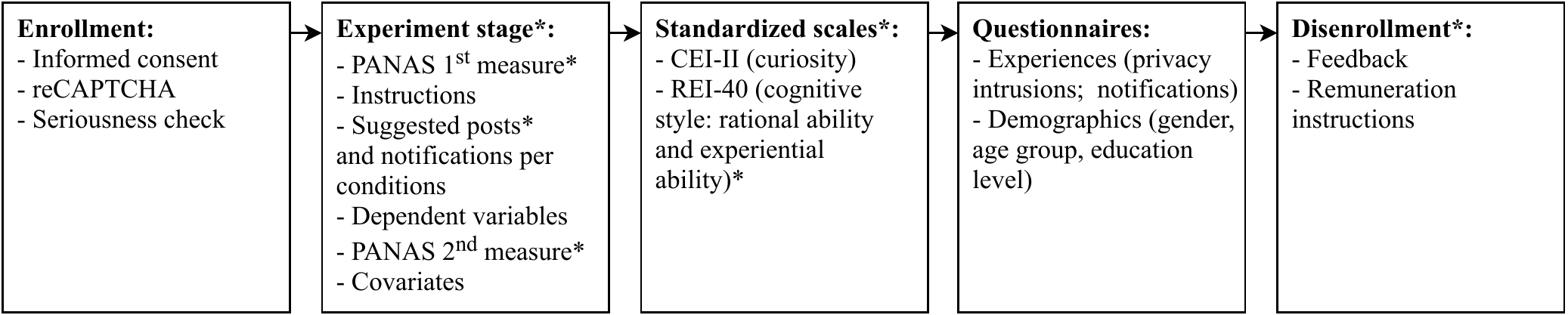}
    \caption{The flow of Experiment 2. * Stages or steps changed, added, or removed compared with the flow of the first experiment.}\label{fig:e2:m:exp_flow}
\end{figure}

The second experiment contained five stages, the flow of which is shown in Fig.~\ref{fig:e2:m:exp_flow}. 

\begin{itemize}[noitemsep, leftmargin=.0cm]
    \item[] \textit{Enrollment:} This stage remained identical to the one in the first experiment.
    \item[] \textit{Experiment:} This stage included the $1^{st}$ PANAS measure, the instruction screen per randomly assigned condition (the suggested post and notifications), the measures of the dependent variables, and the $2^{nd}$ PANAS measure.
    \item[] \textit{Standardized scales:} The participants were asked to answer the CEI-II and, then, a part of the REI-40 scale, measuring \textit{Rational ability} and \textit{Experiential ability}.
    \item[] \textit{Questionnaires:} This stage remained identical to the one in the first experiment, apart from the greater number of age groups available to participants as response options for the sake of convenience.
    \item[] \textit{Disenrollment:} The final screen of the experiment contained an optional feedback form. Upon finishing the experiment, the participants were redirected to collect their remuneration.
\end{itemize}

\subsubsection{Materials}
\label{sec:e2:m:materials}

Overall, the second experiment contained the notifications, suggested posts, two scales measuring individual differences, a set of questions measuring the dependent variables, and a questionnaire of experiences and demographics.

\textit{Suggested posts.} The three suggested posts used in the second experiment differed in the images they contained (Appendix~\ref{sec:app:suggested_post}). 

\textit{Dependent measures: Affective state.} We used the \textit{Positive and Negative Affect Schedule} (PANAS) to measure the valences of participants' positive and negative affective states. PANAS includes 20 items: ten items per both the positive and negative affects.

\textit{Scales measuring individual differences.} We employed two validated scales that assess individual differences:
\begin{itemize}[noitemsep]
    \item CEI-II --- unchanged.
    \item We used the expanded \textit{Rational-Experiential Inventory} (REI-40 \citep{pacini1999REI40}, Appendix~\ref{sec:app:ind_diff:REI40}) to assess cognitive style. We used the part of REI-40 assessing two cognitive styles: conscious, rational cognitive style --- \textit{Rational ability} (10 items); and pre-conscious, experiential style --- \textit{Experiential ability} (10 items).
\end{itemize}

\subsubsection{Participants}
\label{sec:e2:m:participants}

Using the same recruitment procedures as in the first experiment (Section~\ref{sec:e1:m:participants}), and ensuring that the same participants would not be able to participate in the second experiment, we retained $N=765$ complete responses. Table~\ref{tab:e1e2:m:participants} shows the sample demographics.

Regarding the experiences (Appendix~\ref{sec:app:experiences}), $34.9\%$ ($n=267$) of participants recalled experiencing at least one item from the list of online privacy violations ($M=1.55$, $SD=1.80$, range: $0-12$, $N=765$). Regarding familiarity with warnings and notifications, $14.2\%$ ($n=109$) of participants recalled encountering ``in the last month'' at least one of the typical notifications from our list ($M=4.79$, $SD=3.68$, range: $0-15$, $N=765$). Overall, the distribution of participants across the between-subjects groups was balanced: range of $29-35$ participants per each of the 24 independent groups, resulting from the full factorial $2 \times 2 \times 2$ design from Experiment 1 per each of the three suggested posts. The between-subjects groups did not differ in the demographics significantly.
	\subsection{Results}
\label{sec:e2:results}

Prior to the main data analysis, we tested the reliability and validity of the psychometric scales, and examined the correlations.

\subsubsection{Measurement Instruments}
\label{sec:e2:r:meas_ind_diff}

In the second experiment, we used three validated instruments to improve the reliability of the results. We measured the IGPI, using a validated 4-item scale, and the individual differences, using CEI-II for curiosity (containing ten items), and the \textit{Rational ability} and \textit{Experiential ability} sub-scales (each containing ten items) of REI-40 for cognitive style.

\textit{Intention to give personal information.} PCA with orthogonal rotation showed that all 4 items loaded into a single factor, explaining $66.23\%$ of the cumulative variance in the items. Sampling adequacy was acceptable with KMO at~$.66$. Bartlett's test of sphericity was significant at $<.001$. Cronbach's $\alpha=.83$ signified acceptable reliability, which did not increase when removing any of the four items. The IGPI scores were calculated as in Experiment 1 (Section~\ref{sec:e1:r:meas_ind_diff}) as an average of all items.

\textit{Curiosity.} The KMO measure was $.94$, and Bartlett's test of sphericity was significant at $p<.001$, meaning that both the sample and the items were suitable for EFA. This time, all ten items loaded a single factor, explaining $59\%$ of cumulative variance (Cronbach's $\alpha=.92$ for the ten items). The curiosity score was calculated based on ten items, as in the first experiment.

\textit{Cognitive style.} We assessed the participants' cognitive style along two dimensions: \textit{Rational ability} and \textit{Experiential ability}, measured with REI-40. The KMO was $.91$, and Bartlett's test of sphericity was significant at $<.001$. The initial PCA produced three factors. One factor loaded the reversed items, while the other two loaded either positively stated \textit{Experiential style} or positively stated \textit{Rational style} items, respectively. Further reliability analyses, PCAs and the inspection of loading patterns revealed that the sub-scales can be used as intended in the original paper \citep{pacini1999REI40} without removing any of the items. The final calculation of the score for the \textit{Rational style} measure included ten of the \textit{Rational ability} items with the reliability of Cronbach's $\alpha=.83$. The final calculation of the score for the \textit{Experiential style} measure included ten \textit{Experiential ability} items with the reliability of Cronbach's $\alpha=.79$. All items contributed to the reliability of both sub-scales.

\textit{Affective state.} The PANAS measures of the affective states loaded two factors for both instances: taken before and after the manipulation (both times $KMO=.93$, Bartlett's test of sphericity at $p<.001$). The two factors after orthogonal rotation explained $60.86\%$ for the $1^{st}$ PANAS and $65.22\%$ for the $2^{nd}$ PANAS. Both times the first factor loaded all the negative affective state items, and the second one --- all the positive affective state items. The reliability of the measures was excellent: Cronbach's $\alpha=.94$ for the initial negative affect and $\alpha=.91$ for the initial positive affect; Crobach's $\alpha=.94$ for the post-manipulation negative affect and $\alpha=.93$ for the post-manipulation positive affect. All items contributed to the reliability of the measures in all cases. Therefore, four scores were created: \textit{positive affective valence} before the manipulation (PAVb), \textit{negative affective valence} before the manipulation (NAVb), \textit{positive affective valence} after the manipulation (PAVa), and \textit{negative affective valence} after the manipulation (NAVa).

\subsubsection{Descriptive Analysis}
\label{sec:e2:r:descriptives}

Table~\ref{tab:e2:r:corrs} shows relative associations between the measured constructs. There were a few minor differences from the results of the first experiment. This time, we found that curiosity was weakly correlated with both cognitive styles, while the cognitive styles were moderately correlated with each other (which may be explained by the use of a more focused REI version). Additionally, the PAVb and PAVa were both moderately associated with \textit{Curiosity}. 
The affective states before and after the manipulation were strongly correlated with each respective other, while the negative and positive affective states were independent of each other in both instances, indicating that participants' affective states were changing consistently during the experiment. 

\begin{sidewaystable}

    \tbl{Pearson correlations: intention to give personal information (IGPI), preference to restrict the post's visibility (PRPV), preference to confirm posting with involved friends (PCPF), curiosity (CSE), REI-40's rational ability --- rational cognitive style (RCS), REI-40's experiential ability --- experiential cognitive style (ECS), number of recalled privacy violations (NPV), number of recalled warnings and notifications (NWN), PANAS's positive affective valence: before (PAVb), after (PAVa), and their difference (dPAV), PANAS's negative affective valence: before (NAVb), after (NAVa), and their difference (dNAV).}{
\begin{tabular}{llllllllllllll}
         \toprule
         & PRPV & PCPF & CSE & RCS & ECS & NPV & NWN & PAVb & NAVb & PAVa & NAVa & dPAV & dNAV \\
         \midrule
         IGPI & $-.54$*** & $-.42$*** & .34*** & $-.09$* & $-.01$ & .12** & $-.03$* & .16*** & .22*** & .24*** & .17*** & .18*** & $-.09$* \\
         PRPV & 1 & .40*** & $-.28$*** & .03 & $-.02$ & $-.07$ & .02 & $-.08$* & $-.17$*** & $-.15$*** & $-.14$*** & $-.15$*** & .07 \\
         PCPF && 1 & $-.18$*** & .02 & $-.03$ & $-.11$** & .04 & $-.07$ & $-.14$*** & $-.11$** & $-.10$** & $-.08$* & .09*\\
         CSE &&& 1 & .21*** & .16*** & .29*** & .02 & .57*** & .25*** & .57*** & .24*** & .10** & $-.01$ \\
         RCS &&&& 1 & .61** & .13*** & .31** & .21*** & $-.35$*** & .19** & $-.34$*** & $-.02$ & .03 \\
         ECS &&&&& 1 & .09* & .22** & .19*** & $-.26$*** & .17*** & $-.27$*** & $-.01$ & $-.01$ \\
         NPV &&&&&& 1 & .32*** & .12** & .18*** & .11** & .17*** & .01 & $-.03$ \\
         NWN &&&&&&& 1 & $-.02$ & $-.18$*** & $-.03$ & $-.20$*** & $-.02$ & $-.02$ \\
         PAVb &&&&&&&& 1 & .03 & .86*** & .01 & $-.11$** & $-.04$ \\
         NAVb &&&&&&&&& 1 & .05 & .86*** & .03 & $-.28$*** \\
         PAVa &&&&&&&&&& 1 & .07 & .41*** & .04 \\ 
         NAVa &&&&&&&&&&& 1 & .12** & .25*** \\
         dPAV &&&&&&&&&&&& 1 & .17*** \\
         \bottomrule
    \end{tabular}
    }
    \tabnote{***$p<.001$, **$p<.01$, and *$p<.05$. $N=765$.}
    
    \label{tab:e2:r:corrs} 

\end{sidewaystable}

\subsubsection{Effects of Notifications, Posts, and Individual Differences}
\label{sec:e2:r:main_results}

As we study the same RQs (Section~\ref{sec:background}), we performed the analyses on the IGPI, PRPV, and PCPF after ensuring that the data met the necessary assumptions of ANCOVA models. Both the seriousness check and the platform type did not meaningfully alter the models. 

The paired-samples $t$-tests showed that there was a significant difference between the PAVb and PAVa, $t(764)=4.89$, $p<.001$, Cohen's $d=.18$, as well as between NAVb and NAVa, $t(764)=2.69$, $p<.01$, Cohen's $d=.10$. Both times the affective valence decreased after the manipulation: from $M=3.01$, $SD=0.93$ to $M=2.92$, $SD=1.01$ for positive affect, and from $M=1.76$, $SD=0.89$ to $M=1.72$, $SD=0.88$ for negative affect. Therefore, we included the differences between PAVa and PAVb (dPAV), and NAVa and NAVb (dNAV), respectively, as covariates in the models.

\medskip
\textit{Intention to give personal information.} First, we study how the IGPI may be affected by the \textit{Layout}, \textit{Timing}, and \textit{Content} of notifications (first part of RQ1--RQ3), as well as by the individual differences (first part of RQ4). This time, we also include the \textit{Affective states} (first part of RQ5) into the model as covariates, and the \textit{Suggested post variant} as a factor.

We performed an ANCOVA, including the aforementioned factors and covariates, as well as \textit{Curiosity}, \textit{Rational style}, \textit{Experiential style}, the number of recalled notification types, and the number of recalled online privacy violations as covariates (adj. ${R^2=.21}$).

We found a between-subjects main effect of the \textit{Suggested post variant}, $F(2,734)=21.88$, $p<.001$, $\eta^2_p=.06$. Specifically, the \textit{intention to give personal information} was significantly lower among the participants exposed to the \textit{Drinking} post, $M=2.77$, $SE=0.10$, compared with the two new post variants: \textit{Gambling} and \textit{Traveling}, ($M=3.48$, $SE=0.10$, $p<.001$, mean difference $95\%$ CI$[0.38-1.04]$ vs. $M=3.63$, $SE=0.10$, $p<.001$, mean difference $95\%$ CI$[0.53-1.20]$, respectively, with Bonferroni-corrected significance levels). The difference between \textit{Gambling} and \textit{Traveling} on the IGPI was not significant. Next, we did not find the effect of notification \textit{Timing} (RQ2) and \textit{Content} (RQ3), in contrast with the first experiment. However, the effect of \textit{Content} did approach significance, $F(1,734)=3.16$, $p=.076$, $\eta_p^2=.004$: the \textit{Privacy}-related notification still reduced the participants' IGPI, compared with \textit{Neutral} notification of the control group ($M=3.20$, $SE=0.08$ vs. $M=3.40$, $SE=0.08$, respectively). As to the RQ1, we found a null effect of notification \textit{Layout} on the IGPI.

Again, \textit{Curiosity} had a medium-to-large effect on the IGPI, $F(1,734)=95.10$, $p<.001$, $\eta_p^2=.115$ (RQ4), while \textit{Rational style} had a small effect, $F(1,734)=18.85$, $p<.001$, $\eta_p^2=.025$, and \textit{Experiential style} did not have a significant effect (RQ4). The number of notification types the participants recalled and the number of recalled online privacy violations did not appear to explain variance in the IGPI this time. Both the difference in the positive affective valence, and the difference in the negative affective valence significantly inluenced the IGPI, $F(1,734)=20.21$, $p<.001$, $\eta^2_p=.027$ for the dPAV, and $F(1,734)=8.50$, $p<.01$, $\eta^2_p=.011$ for the dNAV.

\medskip
\textit{Preference to restrict the post's visibility.} As before, the participants tended to be more restrictive in terms of visibility of the suggested post for different social groups of the social network (PRPV, $M=4.87$, $SD=1.93$, $Mdn=5.00$, range: $1-7$).

Corroborating the results of the first experiment, the ANCOVA revealed a between-subjects main effect of the \textit{Content} (RQ3), $F(1,734)=7.69$, $p<.01$, $\eta_p^2=.010$, on the \textit{preference to restrict the post's visibility} (adj. $R^2=.11$). A \textit{Privacy}-related message raised the participants' restrictiveness regarding the PRPV, compared with the \textit{Neutral} message ($M=5.06$, $SE=0.09$, $95\%$ CI$[4.88-5.25]$ vs. $M=4.69$, $SE=0.09$, $95\%$ CI$[4.51-4.88]$, respectively; $p<.01$, mean difference $95\%$ CI$[0.11-0.63]$). Additionally, the data demonstrated a significant main effect of the \textit{Suggested post variant}, $F(2,734)=7.38$, $p<.01$, $\eta^2_p=.020$. Participants exposed to the \textit{Drinking} post tended to be significantly more restrictive regarding their post's visibility, than those exposed to the neutral \textit{Traveling} post context ($M=5.19$, $SE=0.11$ vs. $M=4.57$, $SE=0.11$, respectively, with Bonferroni-corrected $p<.001$, mean difference $95\%$ CI$[0.23-1.02]$). The PRPV was not significantly different between the \textit{Drinking} and \textit{Gambling} contexts, or between the \textit{Gambling} and \textit{Traveling} contexts. As to RQ1 and RQ2, we found null effects of \textit{Layout} and \textit{Timing} of notifications on the PRPV.

Addressing the individual differences (RQ4), we yet again found significant effects of \textit{Curiosity}, $F(1,734)=55.74$, $p<.001$, $\eta_p^2=.071$, and of \textit{Rational cognitive style}, $F(1,734)=5.01$, $p<.05$, $\eta_p^2=.007$, and no significant effect of \textit{Experiential style}. This time, the nubmer of recalled notification types and the number of online privacy violations were not significant. The dPAV (positive affect) significantly adjusted the PRPV, $F(1,734)=10.83$, $p<.01$, $\eta^2_p=.015$, as did the dNAV (negative affect), $F(1,734)=4.15$, $p<.05$, $\eta^2_p=.006$ (RQ5).

\medskip
\textit{Preference to confirm posting with involved friends.} The model for the PCPF ($M=3.30$, $SD=1.46$, $Mdn=4.00$, range: $1-5$) included the same factors and covariates as the one described previously (adj. $R^2=.08$). We found a between-subjects main effect of the \textit{Suggested post variant}, $F(2,734)=9.79$, $p<.001$, $\eta^2_p=.026$. The PCPF was significantly higher among the participants exposed to the \textit{Drinking} post, $M=3.68$, $SE=0.09$, compared with the two new post variants: \textit{Gambling} and \textit{Traveling} ($M=3.33$, $SE=0.09$, $p<.05$, mean difference $95\%$ CI$[0.05-0.65]$ and $M=3.13$, $SE=0.09$, $p<.001$, mean difference $95\%$ CI$[0.25-0.85]$, respectively, with Bonferroni-corrected significance levels). The difference between \textit{Gambling} and \textit{Traveling} was not significant.

\begin{figure}
    \centering
    \includegraphics[width=0.8\textwidth]{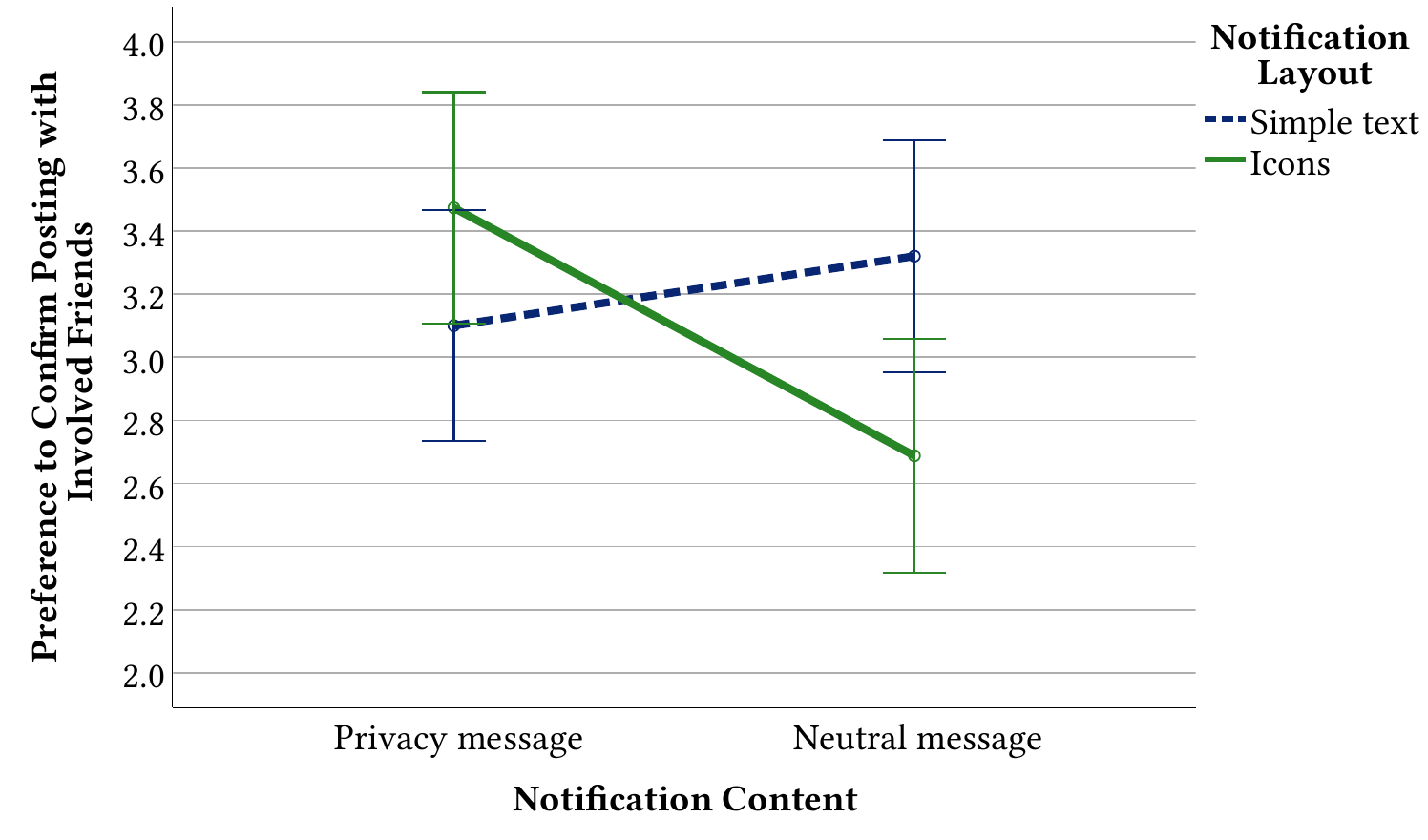}
    \caption{Interaction of notification \textit{Layout} and \textit{Content} within the neutral context of the \textit{Traveling} suggested post. Estimated marginal means, whiskers are $95\%$ CI.}\label{fig:e2:layout:content:interaction}
\end{figure}

We also found a $[\textit{Layout} \times \textit{Content} \times \textit{Suggested post variant}]$ interaction effect, $F(2,734)=6.27$, $p<.01$, $\eta^2_p=.017$, relevant for RQ1 and RQ3. Upon close inspection, the interaction appears within the context of the \textit{Traveling} suggested post (Fig.~\ref{fig:e2:layout:content:interaction}). Specifically, notifications with \textit{Icons} increased the restrictiveness of the participants' PCPF exposed to the \textit{Privacy}-related message, compared with the participants exposed to the \textit{Neutral} content ($M=3.47$, $SE=0.19$ vs. $M=2.69$, $SE=0.19$, respectively), $F(1,240)=18.78$, $p<.01$, $\eta^2_p=.035$. Additionally, the participants who saw the \textit{Neutral} messages tended to be more inclined to confirm with involved friends, when the notification was \textit{Simple text}, compared with \textit{Icons} ($M=3.32$, $SE=0.19$ vs. $M=2.69$, $SE=0.19$, respectively), $F(1,240)=12.25$, $P<.05$, $\eta^2_p=.023$. It indicates how the way the participants responded to the iconized notifications varied for different messages conveyed through such notifications, when participants were seeing a particular \textit{Suggested post variant}. Otherwise, we found a null effect of notification \textit{Timing} on the PCPF (RQ2).

As to the individual differences (RQ4) and affect (RQ5), we found significant effects of \textit{Curiosity}, $F(1,734)=13.75$, $p<.001$, $\eta_p^2=.018$, the number of recalled online privacy violations, $F(1,734)=7.24$, $p<.01$, $\eta_p^2=.010$, and dNAV, $F(1,734)=7.24$, $p<.01$, $\eta_p^2=.010$. Finally, dPAV was approaching significance, $F(1,734)=3.36$, $p=.067$, $\eta_p^2=.005$.

\subsubsection{Individual Differences and Affective States}
\label{sec:e2:r:ind_dif:regres}

Several covariates reflecting the individual differences were significantly related to the IGPI, PRPV, and PCPF. To better understand the relations of these constructs (RQ4 and RQ5), we performed three simultaneous multiple regression analyses --- one for each of the dependent variables. The data were checked for regression assumptions. 

The resulting models (Table~\ref{tab:e2:r:regressions}) were significant: IGPI, $F(7,757)=23.30$, $p<.001$, adjusted $R^2=.17$; PRPV $F(7,757)=12.06$, $p<.001$, adjusted $R^2=.10$; and PCPF $F(7,757)=6.83$, $p<.001$, adjusted $R^2=.05$. \textit{Curiosity} was the strongest predictor in all three models, indicating that an increase in \textit{Curiosity} was associated with an increase in the IGPI, as well as with easing the restrictiveness of both privacy preferences. Additionally, \textit{Rational cognitive style} was associated with a decrease in the IGPI and an increase in the PRPV restrictiveness (the increase in the PCPF was in line with the PRPV, but only approached significance for $\alpha=.05$). Thus, certain individual differences in information processing (RQ4) were correlated with how participants made decisions regarding personal information sharing. \textit{Affective states} (RQ5) were also significantly correlated with all three models, quantitatively corroborating the mixed method findings from the first experiment (Section~\ref{sec:e1:r:qualitative}). Positive affect tended to be associated with an increase in the intention to give personal information and with easing the restrictiveness of both privacy preferences. Negative affect was correlated with a decrease in the intention to share, while stimulating people to be more restrictive with their privacy preferences.

\begin{table}
    \tbl{Joint influence of the curiosity (CSE), rational cognitive style (RCS), experiential cognitive style (ECS), number of recalled privacy violations (NPV), number of recalled warnings and notifications (NWN), difference in positive affective valence (dPAV), and difference in negative affective valence (dNAV) on the intention to give personal information (IGPI) and the privacy preferences (PRPV and PCPF).}
    {
    \begin{tabular}{lrrrrrrrrr}
         \toprule
         Predictor & \multicolumn{3}{c}{IGPI} & \multicolumn{3}{c}{PRPV} & \multicolumn{3}{c}{PCPF} \\
         \cmidrule{2-10}
         & \multicolumn{1}{c}{\textbf{$\beta$}} & \multicolumn{1}{c}{\textbf{$t(757)$}} & \multicolumn{1}{c}{\textbf{$r_p$}} & \multicolumn{1}{c}{\textbf{$\beta$}} & \multicolumn{1}{c}{\textbf{$t(757)$}} & \multicolumn{1}{c}{\textbf{$r_p$}} & \multicolumn{1}{c}{\textbf{$\beta$}} & \multicolumn{1}{c}{\textbf{$t(757)$}} & \multicolumn{1}{c}{\textbf{$r_p$}} \\
         \midrule
         \textbf{CSE} & \textbf{.34***} & \textbf{9.65} & \textbf{.33} & \textbf{$-$.28***} & \textbf{$-$7.44} & \textbf{$-$.26} & \textbf{$-$.15***} & \textbf{$-$3.91} & \textbf{$-$.14} \\
         \textbf{RCS} & \textbf{$-$.19***} & \textbf{$-$4.50} & \textbf{$-$.16} & \textbf{.11*} & \textbf{2.42} & \textbf{.09} & $.08^\dagger$ & 1.76 & .06 \\
         ECS & .05 & 1.27 & .05 & $-.04$ & $-0.81$ & $-.03$ & $-.06$ & $-1.45$ & $-.05$ \\
         \textbf{NPV} & .03 & 0.91 & .03 & .01 & 0.27 & .01 & \textbf{$-$.09*} & \textbf{$-$2.41} & \textbf{$-$.09} \\
         NWN & .00 & $-0.01$ & .00 & $-.01$ & $-0.25$ & $-.01$ & .07 & 1.69 & .06 \\
         \textbf{dPAV} & \textbf{.16***} & \textbf{4.79} & \textbf{.17} & \textbf{$-$.12**} & \textbf{$-$3.48} & \textbf{$-$.13} & \textbf{$-$.08*} & \textbf{$-$2.17} & \textbf{$-$.08} \\
         \textbf{dNAV} & \textbf{$-$.11**} & \textbf{$-$3.24} & \textbf{$-$.12} & \textbf{.08*} & \textbf{2.22} & \textbf{.08} & \textbf{.10**} & \textbf{2.74} & \textbf{.10} \\
         \bottomrule
    \end{tabular}
    }
    \tabnote{***$p<.001$, **$p<.01$, *$p<.05$, and $^\dagger p<0.10$.}
   \label{tab:e2:r:regressions}
\end{table}

\subsection{Discussion}
\label{sec:e2:discussion}

In the second experiment, we improved upon the first one to better investigate the RQs. 

\subsubsection{Notification Properties}
\label{sec:e2:d:notif_propert}

\smallskip
\textit{Layout.} The second experiment showed that when the information is more conventional, notification \textit{Layout} (RQ1) might become helpful for user preference regarding information sharing considering the privacy of others (PCPF).  When the notification contained \textit{Privacy}-related information and the nature of the information was more conventional (\textit{Traveling}, rather than \textit{Gambling} or \textit{Drinking}), the iconized message may have directed participants' attention to the privacy-related implications of their action more than the \textit{Neutral} message containing generic information. On the other hand, when the notification did not alert participants to consider their \textit{Privacy} and that of the others, the iconized notification may have been more visually appealing than the message composed as \textit{Simple text}, resulting in a more relaxed preference to confirm posting with others (this mere suggestion requires further research, perhaps measuring how information layout may affect trust and privacy concerns). This finding ($Timing \times Content \times Suggested{\ }post{\ }variant$) supports our conjecture from the first experiment that participants may be focusing their attention on the available layouts, depending on the specific context of the interaction, i.e., the suggested post variant in our case. The second experiment did not reveal additional effects of the notification \textit{Layout} on the IGPI and PRPV.

\smallskip
\textit{Timing.} Contrary to the first experiment, we did not confirm the effect of \textit{Timing} (RQ2) of notifications on the IGPI. There was again no such effect for the PRPV and PCPF. These findings suggest that the extent to which feedback timing helps make information more relevant and informs users' decisions may be limited and situational.

\smallskip
\textit{Content.} The second experiment confirmed the effect of notification \textit{Content} (RQ3) for the preference regarding with whom to share information (PRPV). However, it did not corroborate the same effect for the intention to share (IGPI --- though it approached significance at $p<.10$). Based on one robust effect on the PRPV, a $[Layout \times Content \times Suggested{\ }post{\ }variant]$ interaction affecting PCPF (and a borderline significant effect on the IGPI), we argue that the notifications were not discarded as mere interruptions, and feedback may help to inform users and to allow them to control their privacy and self-presentation.

\subsubsection{Individual Differences in Information Processing}
\label{sec:e2:d:ind_diff}

The significant effects of \textit{Curiosity} and \textit{Cognitive style} for decisions regarding the intention to give personal information and privacy preferences were in line with the results of the first experiment (RQ4).

Assuming that \textit{Curiosity} reflects a willingness to seek and embrace new information and experiences, it may lead to more relaxed and open behavior. This may be part of an explanation of how curiosity motivates the intention to share (IGPI) and reduces the tendency to restrict information spread (PRPV), including considerations for the privacy of others (PCPF).

The \textit{Rational cognitive style} (\textit{Rational ability}), being analytical, information-focused, may have helped participants pay closer attention to the information provided in the notifications and suggested posts and consider the privacy implications of their actions. This can reasonably explain how rational cognition acted in the opposite direction to curiosity for both the intention to give personal information and the preference to restrict a post's visibility (also, in Experiment 2, for the preference to confirm posting with involved friends, albeit marginally). Simultaneously, the \textit{Experiential cognitive style} (\textit{Experiential ability}) did not play a role in how people decided upon their personal information sharing attitudes. On the one hand, the mere fact of participating in the experiment may have caused participants to be more deliberate and focused, limiting or subduing their preconscious cognition to some degree. On the other hand, experiential cognition may simply not be relevant in the interactions involving information sharing.

\subsubsection{Affect}
\label{sec:e2:d:affect}

The second experiment replicated the previous findings regarding the ubiquitous role of affective states in personal information sharing (within the boundaries of how sharing was operationalized in our experiments, RQ5). Having measured positive and negative affect separately, we showed how the two affective dimensions might act on the sharing attitudes in opposite directions, in line with the \textit{affect-as-information} and \textit{feeling-as-information} theories \citep{Schwarz2012,Schwarz2007,Clore2001}. The overall emotional state, resulting from the balance of positive and negative affect, may be an important factor for people's attitudes regarding personal information sharing in a given moment. 

\subsubsection{Suggested Post Variants}
\label{sec:e2:d:sugg_post_var}

The two additional suggested posts we added were meant to broaden the scope of the first experiment and to see how its results would fare with other content, where either the nature of the depicted experience is different, or the information is less sensitive. Indeed, the suggested post proved to be the strongest contextual cue, influencing ensuing personal information sharing attitudes. The original \textit{Drinking} context appeared to be a much stronger stimulus, leading to overall more cautious attitudes (IGPI, PRPV, and PCPF). It also seems that the visual design of privacy notifications matters more when the shared information is more conventional (\textit{Traveling}). The absence of other interactions with the suggested post variants indicates that the main results are generalizable.

	\section{General Discussion}
\label{sec:general_discussion}

Our two online experiments dealt with assisting users' decisions and improving their informedness about the implications of privacy-related actions. Specifically, we looked at how certain individual differences\footnote{Even though around a third of participants in each experiment reported to be 65 or older, the experiments used bespoke dependent variables and were not designed to test the effects of the demographics, such as differences across generations or by age. Thus, the study did not aim to have equally representative samples within each age cohort. Additionally, there was too large a difference in the size of the age cohorts to include the age into the analysis without grouping up the participants (e.g., median split, quartile split).} and properties of notifications may affect the sharing of personal information and privacy preferences. The results, summarized in Table~\ref{tab:gd:summary}, provide us with insights and have implications for the five research questions we posed in Section~\ref{sec:background}.

\begin{table}
    \tbl{Summary of the effects of the notification properties, suggested posts, and individual differences on the information sharing attitudes in the two experiments.}
    {
    \begin{tabular}{llllllll}
         \toprule
         \multicolumn{1}{c}{Variable,} & Level/Type & \multicolumn{2}{c}{\textit{Intention to give}} & \multicolumn{2}{c}{\textit{Preference to re-}} & \multicolumn{2}{c}{\textit{Preference to con-}} \\
         \multicolumn{1}{c}{\textit{statistic}} && \multicolumn{2}{c}{\textit{personal infor-}} & \multicolumn{2}{c}{\textit{strict the post's}} & \multicolumn{2}{c}{\textit{firm posting with}} \\
         && \multicolumn{2}{c}{\textit{mation}} & \multicolumn{2}{c}{\textit{visibility}} & \multicolumn{2}{c}{\textit{involved friends}} \\
         \cmidrule{3-8}
         && \multicolumn{1}{c}{Exp 1} & \multicolumn{1}{c}{Exp 2} & \multicolumn{1}{c}{Exp 1} & \multicolumn{1}{c}{Exp 2} & \multicolumn{1}{c}{Exp 1} & \multicolumn{1}{c}{Exp 2} \\
         \midrule
         Layout, $M$ & \textit{Simple text} & 2.36 & 3.33 & 5.58 & 4.83 & 3.99 & 3.35 $^\times$ \\
         & \textit{Icons} & 2.54 & 3.26 & 5.67 & 4.92 & 3.90 & 3.41 $^\times$ \\
         
         Timing, $M$ & \textit{Before} & 2.30* & 3.31 & 5.70 & 4.90 & 3.98 & 3.38 \\
         & \textit{After} & 2.60* & 3.28 & 5.55 & 4.86 & 3.91 & 3.37 \\
         
         Content, $M$ & \textit{Privacy} & 2.30* & $3.20^\dagger$ & 5.77* & 5.06** & 4.02 & 3.43 $^\times$ \\
         & \textit{Neutral} & 2.60* & $3.40^\dagger$ & 5.49* & 4.69** & 3.87 & 3.32 $^\times$ \\
         \midrule
         Suggested & \textit{Drinking} && $2.77^A$ && $5.19^A$ && $3.68^A$ \\
         \hspace{3ex}post, $M$ & \textit{Gambling} & n/a & $3.48^B$ & n/a & $4.87^{AB}$ & n/a & $3.33^{B}$ \\
         & \textit{Traveling} && $3.63^B$ && $4.57^B$ && $3.13^B$ $^\times$ \\
         \midrule
         Curiosity, $r_p$ && .34*** & $.33$*** & $-.20$*** & $-.26$*** & $-.11$* & $-.14$*** \\[1ex]
         
         Cognitive & \textit{Rational} & $-.13$** & $-.16$*** & .16*** & $.09$* & $.10$* &  $.06^\dagger$ \\
         \hspace{3ex}style, $r_p$ & \textit{Experiential} & $.00$ & .05 & $-.03$ & $-.03$ & $-.05$ & $-.05$ \\
         \midrule
         Affective state, & \textit{Positive} & $3.93^A$ & $.17$*** & $4.45^A$ & $-.13$** & $3.23^A$ & $-.08$* \\
         \hspace{3ex}$M$ for Exp 1, & \textit{Neutral} & $2.73^B$ & n/a & $5.25^B$ & n/a & $3.81^B$ & n/a \\
         \hspace{3ex}$r_p$ for Exp 2 & \textit{Negative} & $1.98^C$ & $-.12$** & $6.03^C$ & $.08$* & $4.20^B$ & $.10$** \\
         \bottomrule
    \end{tabular}
    }
    \tabnote{\textit{Note}: $M$ --- estimated marginal means (main effects), $r_p$ --- partial correlations, n/a --- not applicable;
    
    For 2-level variables and covariates: ***$p<.001$, **$p<.01$, *$p<.05$, $^\dagger p<.10$;
    
    For 3-level variables: A, B, C --- different letters denote significantly different estimated marginal means; 
    
    $^\times$variables in the three-way interaction, estimated marginal means not shown.}
   \label{tab:gd:summary}
\end{table}

\subsection{Notification Properties}
\label{sec:gd:notif_propert}

\subsubsection{Layout}
\label{sec:gd:layout}

There was no robust evidence for an effect of the information layout (\textit{Icons} vs. \textit{Simple text}) on people's privacy attitudes in our experiments (RQ1), despite the arguments made in \citet{waddell2016paraphraseddesign}, \citet{tabassum2018incrusattcomicpolicy}, and \citet{Angulo2012tousapripodispman}. The information layout only mattered for one of the privacy preferences in the second experiment as part of a three-way interaction with notification content when the shared information was hardly sensitive. These results indicate that the situation context may be a prominent factor defining whether or how the visual layout of notifications plays a role in helping people attend to notification content. System designers should perhaps structure and iconize information aids when they expect the users to be less alert or concerned. This may help highlight the implications, risks, or actions needed in a given situation, assuming the situation might not provide a salient enough reason for the users to be motivated or interested in the moment. Nevertheless, more research on the effects of notification layout and content in different contexts is needed.

\subsubsection{Timing}
\label{sec:gd:timing}

Even though notification \textit{Timing} (RQ2) did not have a consistent effect in the two experiments, we would still advise system designers and privacy practitioners to inform users \textit{Before} they commit to decisions. Feedback timing may help decision-making regarding information sharing, preventing regrettable disclosures (Experiment 1). Yet, the effect of timing can be circumstantial, as information sharing appears to be consistently influenced by such situational and momentary factors as sharing context, affective states, and the cognitive effects (curiosity, thinking style), but not by the notification timing (Experiment 2).

\subsubsection{Content}
\label{sec:gd:content}

In our experiments, we used context-relevant and irrelevant message \textit{Content} of notifications to control for the mere interruptiveness of notifications (RQ3). Overall, the notifications were systematically acknowledged, even though less so in the second experiment. Such results imply that users may pay attention to the content of messages they receive (at least when the message is concise), and it is not the mere interruption that affects people's choices or makes them adjust their privacy attitudes (IGPI, but especially PRPV). Indeed, the feedback content can actively help users avoid regrettable choices in terms of consequences for their image, reputation, and privacy. Yet, it is challenging to keep content relevant, since system designers would have to make sure that their systems can identify when users are least busy and most interested in context-dependent feedback.

\medskip

The effects of notification properties should not be considered without regarding such a factor as notification fatigue: when users are exposed to an excessive amount of (not always necessary or timely) notifications, they may become apathetic and disregard further notifications altogether. However, even when users think that notifications are a nuisance, the notifications still remain a means to inform the users. Notifications can be personalized, can present information in a timely and concise format, and simplify, prioritize, and structure the information. Studying notification properties should allow researchers and system designers to make notifications more versatile and relevant, minimizing unnecessary user exposure to notifications. Therefore, more research is needed to determine whether and how the notification layout and timing (among other properties) may help alleviate users' fatigue and allow users to attend to the content of the notifications more carefully.

\subsection{Individual Differences in Information Processing}
\label{sec:gd:ind_diff}

Individual differences in how people process new information (RQ4) --- \textit{Curiosity} and \textit{Cognitive style} --- had a robust effect on  people's privacy attitudes, both behavioral intention (IGPI) and privacy preferences (PRPV and PCPF).

Curiosity appears to be a crucial trait, consistently affecting privacy attitudes in both experiments. It appears that curiosity can be instrumental in aiding user decision-making, possibly by increasing their tendency to seek new information. System designers should be interested in encouraging or supporting curiosity in users, because curiosity may motivate the users to learn about or accept the consequences of their actions. However, we also caution that the higher levels of curiosity during a given online interaction may make users more vulnerable to manipulation and expose them to risks of over-disclosure and regrettable actions regarding their information sharing. This extends the literature on curiosity and regret aversion in decision-making \citep{VANDIJK2007curikilregre} to decisions with privacy implications, such as personal information sharing on social networks. When user actions may entail substantial consequences to the users' self-presentation and privacy, system designers and privacy practitioners should not lure users into over-disclosure --- not letting ``curiosity kill the cat''. As to the methods, the researchers may need to investigate further how curiosity can be induced or subdued. More research is needed on the effects of curiosity on privacy-related behaviors.

The \textit{Rational cognitive style} (i.e., Need for Cognition in REI-10, or \textit{Rational ability} from REI-40) was related to the decision-making processes regarding personal information sharing. Rationality enables an analytical approach, aiding information seeking and leading to better informed decisions regarding personal information sharing. System designers, regulators, and privacy practitioners should encourage rational human cognition to support users' decision-making regarding privacy. This can be done by ensuring that all the needed information is accessible in an intelligible format (for instance, using comic-based designs \citep{tabassum2018incrusattcomicpolicy} or dejargonized phrasing and uncluttered layout \citep{waddell2016paraphraseddesign}). Lack of evidence for the role of the \textit{Experiential cognitive style} should not discourage further research. Perhaps experiential cognition might be more relevant for experience-based, automatic, learned processes (like in the case of previously discussed car driving applications in \citet{kehr2015thinkstylesprideci}. Researchers may want to direct more attention to the potential effects of cognition in privacy-related interactions.

In a broader sense, both curiosity and cognitive styles should also be considered in systems design. As curiosity motivates information search and learning, while rational or experiential thinking governs how the information is being processed, researchers and practitioners in human-computer interaction (HCI) should consider these individual differences for user interface design and user experience. Supporting curiosity and, perhaps, experiential cognition (of course, given there is little negative consequence for user privacy) may be instrumental in increasing user engagement, as well as user satisfaction. Providing for rational cognition may help users better understand systems and avoid unwanted outcomes, especially when encouraging attention and caution is a priority. Thus, curiosity and cognitive styles may become effective factors of system customization, providing different people with tools to achieve their practical goals in using the systems in the ways they find individually convenient, according to their personal predispositions. Additionally, curiosity and cognitive styles may be considered as selection criteria in certain areas of employment, in professional contexts, when expertise and specific skills are desirable and may be related to these individual differences. Crucially, more research is needed to understand how to encourage each of these individual differences independently or in combinations, as well as to establish their effects in general and professional contexts broader than, but still inclusive of, privacy-related decision-making.

\subsection{Affect}
\label{sec:gd:affect}

Affective states (RQ5) were consistently related to privacy attitudes (IGPI, PRPV, and PCPF) in our experiments. People in more positive affective states may pay less attention to external information (in line with \citet{Schwarz2012}, \citet{Schwarz2007}, and \citet{Clore2001}), which can make them more vulnerable to manipulations regarding their personal information sharing and disclosures. Our results corroborated the findings of \citet{Coopamootoo2017} and \citet{kitkowska2020epttvdopnetioccaa} who showed increased sharing attitudes with positive affect and increased protective attitudes with negative affect. 

Our results also highlighted the role of the affective state for PCPF, which reflects consideration for the privacy of others. Prior literature showed that some privacy considerations, when sharing information about others, might be important for the use of privacy controls \citep[for instance, perceived shared risks,][]{james2017expothersosn}. We argue that the affective states may be another antecedent of privacy considerations for others. More attention should be given to the effects of emotions in privacy decision-making to help people avoid dealing with the consequences of over-disclosure or regrettable disclosures. System designers should be aware of the potential risks of affective designs aimed at increasing positive affect.

\subsection{Note on Effect Sizes}
\label{sec:gd:NoteOnEffectSizes}

We found multiple significant effects, most of which may be considered small to medium in size ($\eta^2_p$, $\eta^2$, and $\omega^2$ in the range of $.01-.12$), according to a common interpretation \citep[e.g.,][]{maher2013EffectSize} developed from Cohen's work. This means that the results reported in this paper should be interpreted with caution. However, the Cohen's convention regarding the effect sizes is only a general recommendation and lacks consensus, especially regarding models with covariates in research related to psychology. \citet{Schafer2019} show that for between-subjects designs, the effects tend to be smaller than for within-subjects designs. \citet{Lakens2013} argues that in such cases, the effect sizes should not be interpreted based on the Cohen's convention, while \citet{Baguley2009StandOrSimpleEffSize} argue that the standardized effect sizes should not be reported at all, favoring simple effect sizes instead. Nevertheless, it is important to keep proportionality in mind when interpreting effect sizes: a statistically small effect translates to a large practical effect, when the studied population is large. Given that the experiments were studying online sharing decisions of a general adult audience (users of online social networks), the notification properties and individual differences may affect the sharing preferences of a large number of users in practice. Additionally, to reach a better understanding of the effects, it is important to consider and compare them in replication studies, which should be kept in mind for future research.

\subsection{Contributions and Implications for Research}

We obtained empirical evidence showing that the timing of feedback may affect the intention to give personal information in the context of sharing on social networks, albeit circumstantially. Notifications informing users before they make decisions may result in lower sharing intentions. Further, we demonstrated that feedback with context-dependent meaningful content is distinguishable by users from identically interruptive but context-irrelevant feedback. The relevance of feedback may be one of the crucial parameters defining feedback effectiveness and efficiency in aiding users' informed decision-making. Our results also hinted that differences in visual designs alone might not be enough to affect users' sharing attitudes and privacy preferences, and might depend on the situational context or the nature of the information being shared, and on the notification content. Indeed, contextuality may be the key: even though the suggested posts were added to broaden the scope the findings, the revealed effects provide empirical support to the argument that users' privacy-related decision-making may rely on norms inferred from contextual information \citep[in line with][]{Nissenbaum2004privacycontintegr}. The effects of the suggested posts relate to the effect the context may have on ``privacy expectations''~\citep{martin2016measuring}, affirming that contextual cues may influence information sharing attitudes. However, as the suggested posts were not developed from a research question or theory a priori, we did not include manipulation check in the experiment, and should be cautious with conclusions we draw regarding their effects. The original suggested post from Experiment 1 does appear to be more sensitive or controversial, given that the main differences were found between it and the two additional posts in Experiment 2. Participants were more cautious --- less open to share and more restrictive in terms of privacy preferences --- when exposed to the original post. Simultaneously, the combination of the notification layout and content was effective, when the post referred to traveling and sightseeing, which may have been perceived as the most conventional and inoffensive. These findings indicate that responses to notifications may in part rely on the nature of information being shared (in addition to the notification properties and individual differences). This factor can be crucial for notifications' usability and effectiveness, and it should be carefully considered and systematically addressed in future research.

We explored curiosity and cognitive style and collected empirical evidence that such individual differences should be systematically addressed in the study of privacy decision-making and should not be neglected in practical applications. These individual differences in information processing may be responsible for modulating information search and comprehension, aiding users' informed decisions, and adjusting the effectiveness of notifications. Considering cognitive style in the study of privacy decision-making may provide more insight on how to improve user experience with, and usability of system customization (with cognitive-experiential self-theory \citep{Epstein2012CEST} being one approach to relate cognitive styles to systems' us). Curiosity is a trait necessary for exploratory behavior; in particular, it can motivate such behavior~\citep{Litman2005}. Its intrinsic motivational characteristics reflect in a drive to seek meaningful interests and desires~\citep{Kashdan2004}. Some researchers proposed that curiosity has two different components: diversive and specific~\citep{Kashdan2004}. The former is more explorative and involves scanning, recognizing, and affecting attention given to novel experiences. The latter is tied to the specific activity and engagement in such an activity, resulting in pleasure, discovery, or use of skills. Only when both components of curiosity are present, then curiosity may result in learning. The notifications and the task used in the experiments might have been familiar enough to the participants (well-known context of sharing on social networks, pop-up notifications), so that curiosity measured in the experiments might have relied on the diversive rather than both components. Hence, the aspect of learning is out of the scope of the paper. Additionally, curiosity might lead to exploratory behavior when information is either unknown or partially known \citep{Litman2005}, yet the potentially familiar setting in the experiments might have elicited participants' feeling of ``known''. Therefore, incorporating the aspects of learning and exploration should be considered in future research. 

Finally, we found that the affective states experienced at a given moment can influence users' attitudes and preferences. The affective states can change reliance on external information, regulating the attention users give to privacy notifications. The study of emotions in privacy decision-making \citep[relying, for instance, on affect-as-information or feeling-as-information theories,][]{Clore2001,Schwarz2007,Schwarz2012} may provide promising insight into how to structure online interactions without hindering user autonomy or causing regret.

\subsection{Limitations and Future Work}
\label{sec:gd:lim_fw}

The experiments relied on vignettes and self-reports, possibly limiting the ecological validity of the results. However, the experimental task required little effort and time, and included non-restrictive attention and seriousness checks. Moreover, the participants' feedback and responses to the open-ended question give us confidence in the effectiveness of the manipulation and the reliability of the results. Also, the external validity may be limited, as the experiments were restricted to English-speaking residents of the United States. Nevertheless, we observed more than satisfactory reliability of measurements and critical test values, which can inform further investigations on privacy attitudes and behaviors. Additionally, the amount of text in the privacy vs. neutral notifications differed. To mitigate this, we kept the number of paragraphs and (or) lines identical in the notifications' layout and kept the differences in the readability and lexical density as minor as possible.

Overall, the data did not consistently show the two-way interactions of $[Timing \times Content]$. This implies that these properties of notifications had an independent impact, making space for future research in contexts not limited to sharing on social networks. Future research can also expand the combinations of notification properties to include other possible types of notification timing (e.g., time delays), content (e.g., framed messages), and other factors. Crucially, privacy decision-making research should account for the individual differences in human cognition, such as curiosity and cognitive style. The study of the emotions in privacy decision-making is another potential avenue for research. 

Next, even though we did not include the nature of shared information (suggested posts) as another research question in this paper (and, therefore, did not have the necessary manipulation check to draw definitive conclusions), the results showed that the nature of shared information is a significant factor, which should be considered in future studies. Also, we did not collect accurate data to assess and verify which particular devices participants used as it was not a research question in the paper and we were following the best practices in data protection, including but not limited to the principle of data minimization. Adjusting the methodology to include such usability-related research questions can be considered in future studies. Additionally, as the age, gender, and education may be related to certain general privacy attitudes, a systematic study of the demographic effects can be included in future research. Finally, future research may extend the study of the influence of the aforementioned factors to other privacy attitudes and perceptions, as well as to the study of associated behaviors.

	\section{Conclusion}
\label{sec:conclusion}

Sharing personal information online entails consequences for the users' privacy and self-presentation, sometimes resulting in regrettable disclosures. Assuming that the information sharing attitudes are ``attitudes toward behaviors'', the most prominent factor impacting the sharing attitudes is the information being shared. Nevertheless, we have demonstrated that notification properties influence such attitudes as the intention to share (give personal information) and the privacy preferences motivated by self-presentation (restricting visibility of posts) or by considerations about others (confirming posting with friends). Overall, privacy-related notifications can better inform users' decisions regarding sharing when the notifications appear before users commit to their decisions and the information layout is structured (especially, when the context of sharing is more conventional). The individual differences in information processing --- curiosity and rational cognition --- are unequivocally related to the sharing attitudes. Generally, curiosity is associated with reduced privacy considerations, while rationality is associated with increased privacy considerations. Moreover, positive and negative emotions consistently relate to the sharing attitudes in a similar pattern: positive affect is associated with reduced considerations for privacy, whereas negative affect is associated with heightened considerations for privacy. Future research should systematically address the effects of cognitive styles, curiosity, and affective states on user information sharing and other privacy-related attitudes and behaviors. The development of a mechanism accounting for such individual differences to make the design, timing, and content of privacy-related notifications context-dependent can be a viable direction for future research.

	\section*{Funding}
		This research is partially funded by the EU Horizon 2020 research and innovation programme under the Marie Sk{\l}odowska-Curie grant agreement No 675730 ``Privacy and Us''. The funding source had no involvement in study design; the collection, analysis, and interpretation of data; the writing of the report; and in the decision to submit the article for publication.
		
	\section*{Disclosure statement}
		The authors confirm that there are no relevant financial or non-financial competing interests to report.
	
	\bibliographystyle{apacite} 
	\bibliography{main}

	\pagebreak
	\section*{About the Authors}

	    \noindent\textbf{Yefim Shulman} received a Master’s degree in business informatics from the Higher School of Economics, Moscow, Russia. He is pursuing a Ph.D. at Tel Aviv University in Israel. Yefim’s research deals with usable privacy and human-computer interaction, focusing on how to inform user decision-making regarding online privacy and self-presentation.

	\bigskip

        \noindent \textbf{Agnieszka Kitkowska} holds an MA in History of Art and Culture from the University of Nicolaus Copernicus in Toru\'{n}, Poland, an MSc in Computing from Edinburgh Napier University, Scotland, and a Ph.D. in Computer Science from Karlstad University. Her work focuses on privacy-related decision-making, HCI, and technology and behavioral change. 
        
    \bigskip

        \noindent\textbf{Joachim Meyer} is the Celia and Marcos Maus Professor for Data Sciences in the Department of Industrial Engineering at Tel Aviv University. He holds a Ph.D. (1994) in Industrial Engineering from Ben-Gurion University of the Negev, Israel. He is an elected fellow of the Human Factors and Ergonomics Society.	
	
	\appendix
\newpage
\section{Notification Designs Used in the Experiments}
\label{sec:app:notificitions}

\begin{figure}[ht!]
\centering
\subfloat[A1C1]{\includegraphics[width=.49\textwidth]{Figures/A1C1.pdf}}\hspace{5pt}
\subfloat[A1C1m]{\includegraphics[width=.27\textwidth]{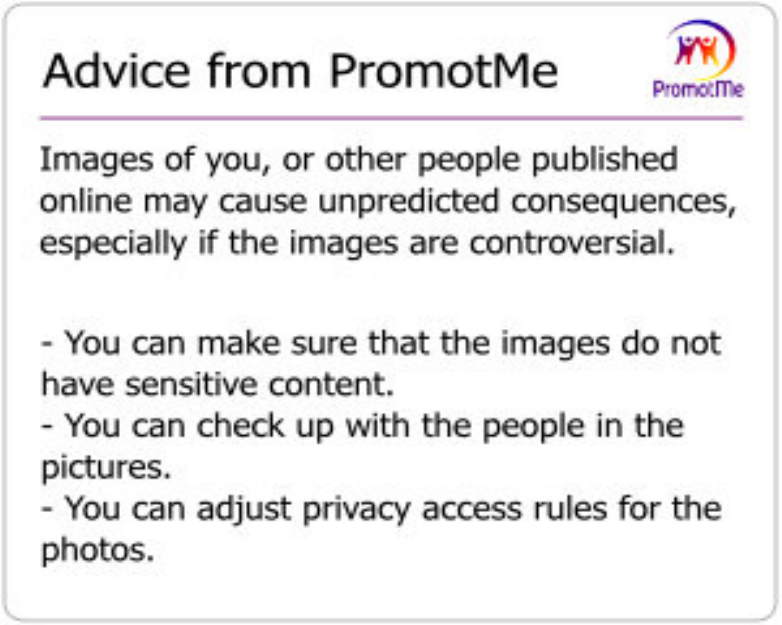}}\hspace{5pt}

\subfloat[A1C2]{\includegraphics[width=.49\textwidth]{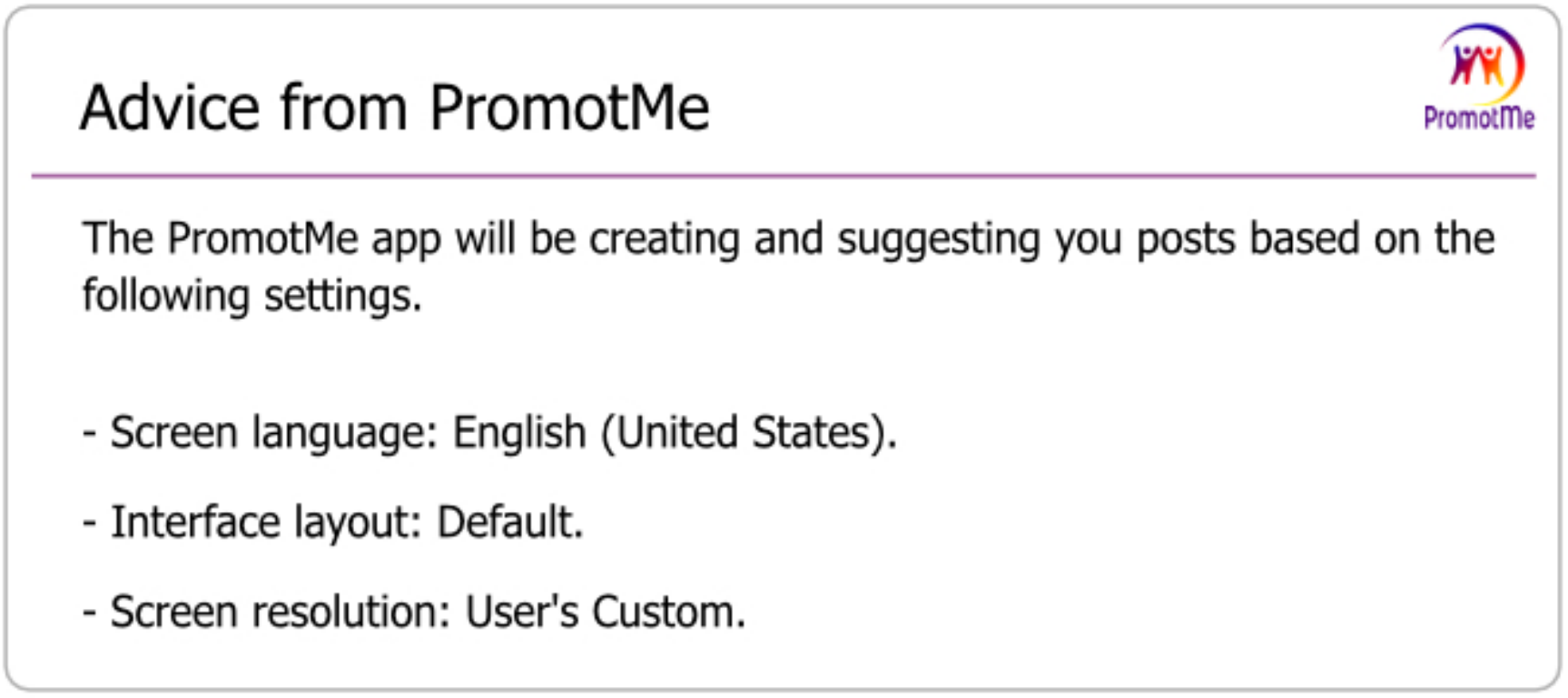}}\hspace{5pt}
\subfloat[A1C2m]{\includegraphics[width=.27\textwidth]{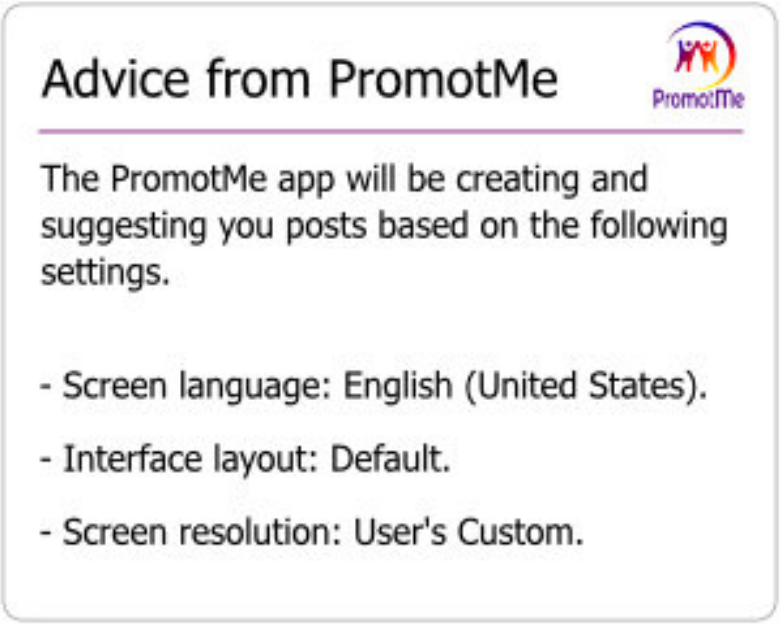}}\hspace{5pt}

\subfloat[A2C1]{\includegraphics[width=.49\textwidth]{Figures/A2C1.pdf}}\hspace{5pt}
\subfloat[A2C1m]{\includegraphics[width=.16\textwidth]{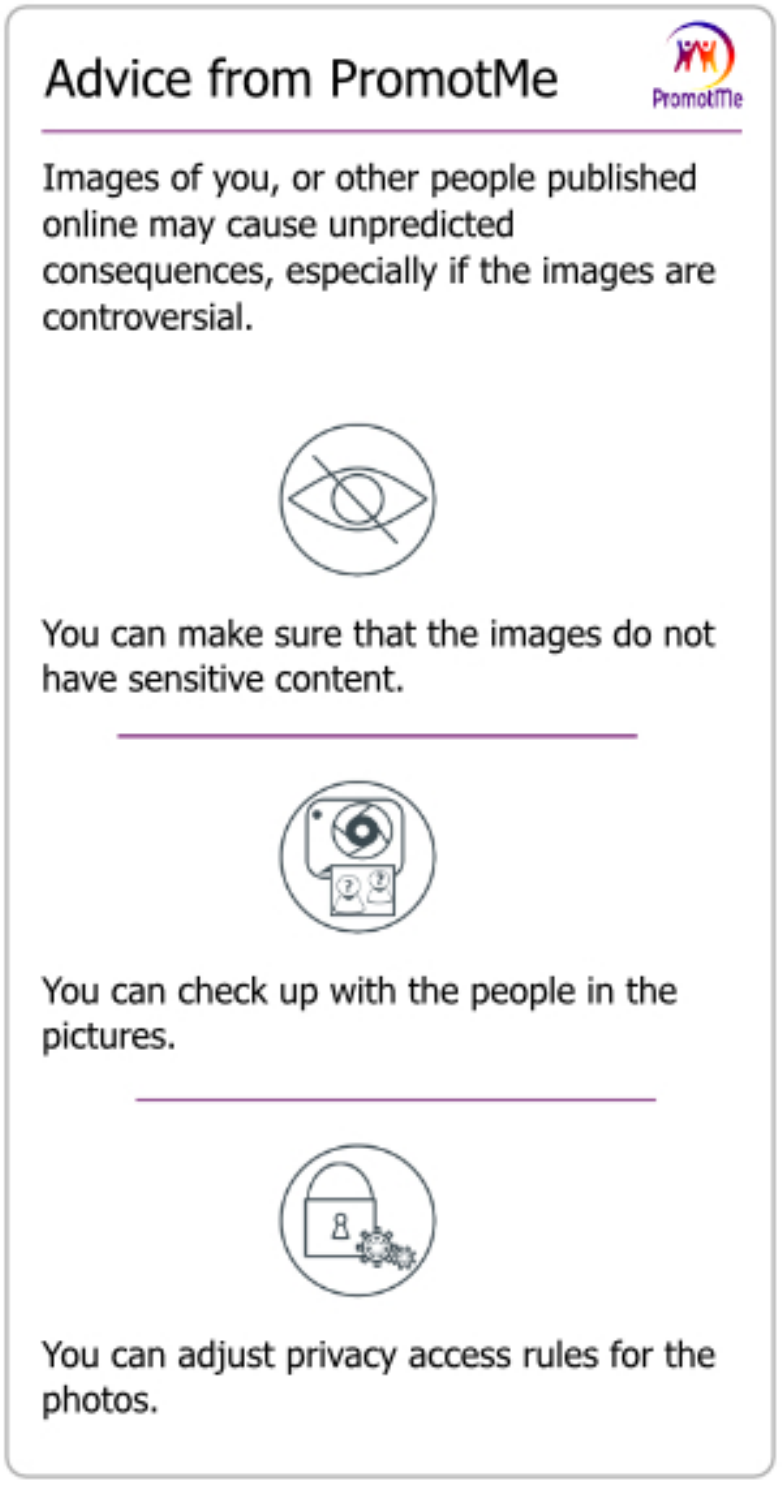}}\hspace{5pt}

\subfloat[A2C2]{\includegraphics[width=.49\textwidth]{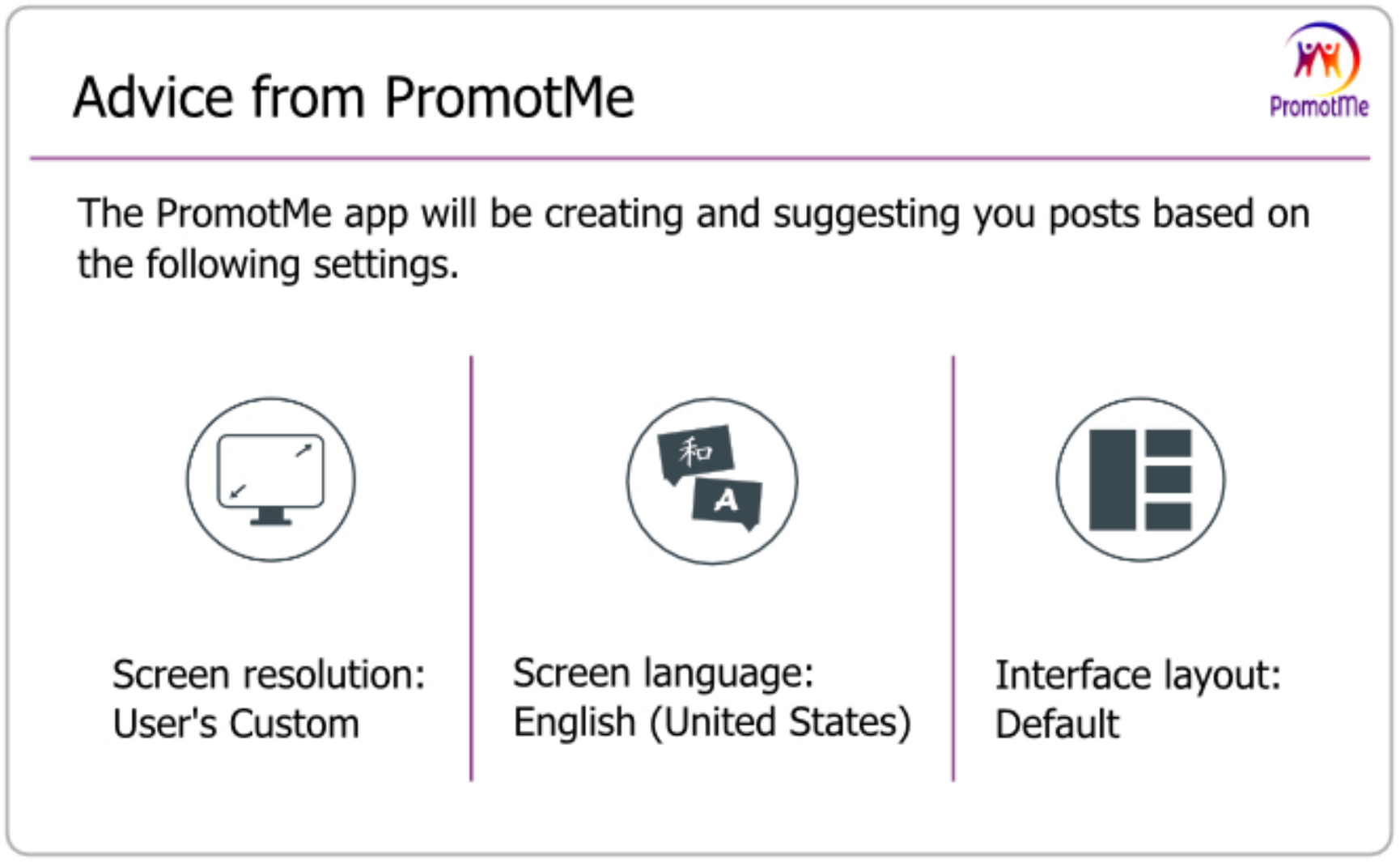}}\hspace{5pt}
\subfloat[A2C2m]{\includegraphics[width=.16\textwidth]{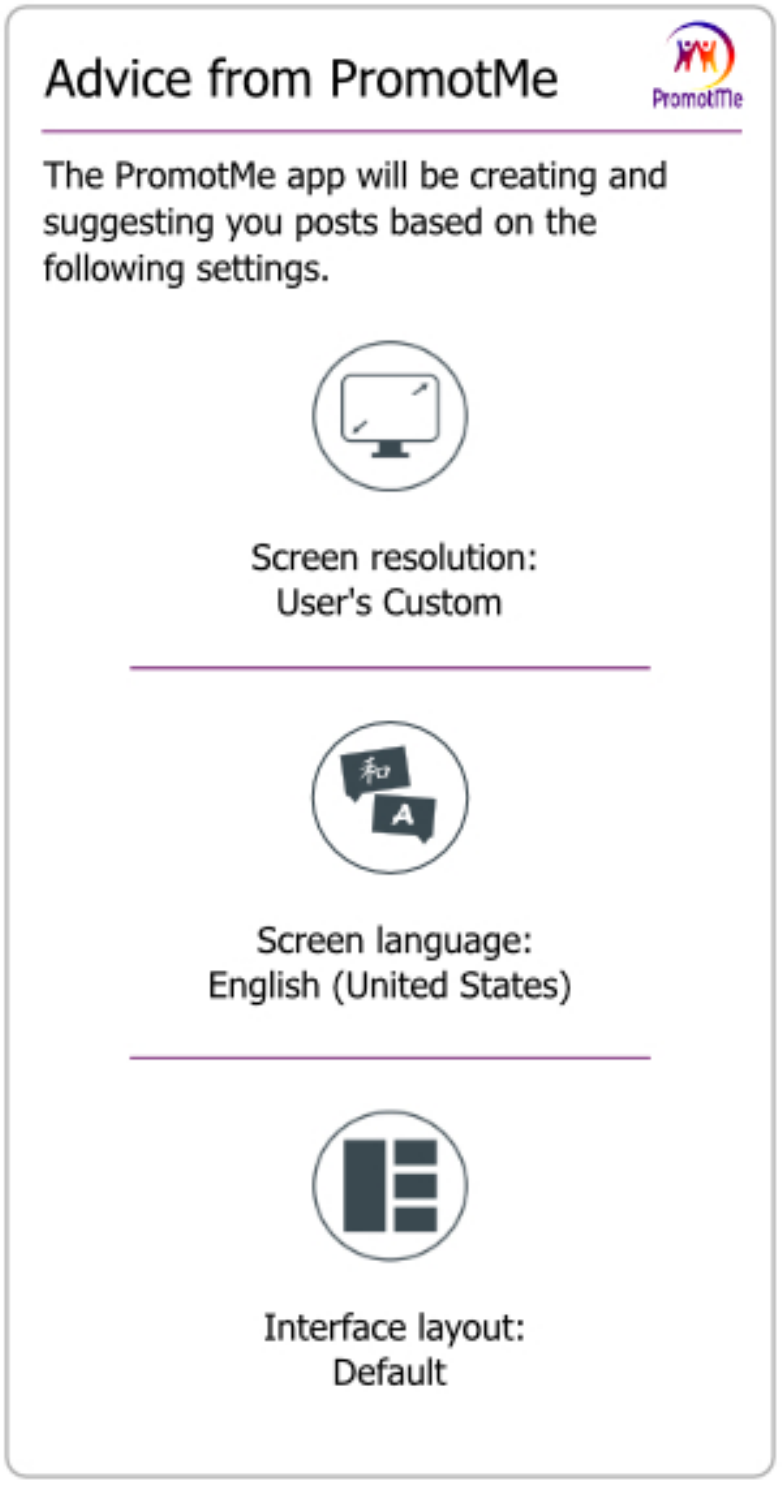}}\hspace{5pt}
\caption{Design of notifications used in the experiments on web (-) and mobile (m) platforms. The alphanumerical labels correspond to the levels of the independent variables: A1C1 --- indication with \textit{Simple text} and \textit{Privacy} message fitted for the web; A2C1m --- indication with \textit{Icons} and \textit{Privacy} message fitted for mobile, etc.}
\label{fig:app:notifications}
\end{figure}

\pagebreak

\section{Measures of Dependent Variables}
\label{sec:app:DVs}

\subsection{Intention to Give Personal Information}

Four items adapted from \citet{malhotra2004iuipc} with minor modification to anchors. The scale reliability was reported in the original paper as Composite Reliability (CR) and Average Variance Extracted (AVE) at the levels of $CR(IGPI)=.95$ and $AVE(IGPI)=.86$~\citep{malhotra2004iuipc}.

\subsubsection*{Participant instructions:}
``Think about publishing this post in reality. How strongly would you agree or disagree with the following paired statements?

Please, indicate your opinion by choosing and clicking a dot between the two options in each row. The closer the dot is to either side of each row, the more you agree with that side of the statement.''

\subsubsection*{Rating scale and anchoring}
Seven-point semantic differential rating scale anchored with paired statements:
\begin{enumerate}[noitemsep]
    \item I'm unlikely to publish -- I'm likely to publish
    \item For me, publishing is not probable -- For me, publishing is probable
    \item For me, publishing is possible -- For me, publishing is not possible [Anchoring reversed]
    \item I'm willing to publish -- I'm unwilling to publish [Anchoring reversed]
\end{enumerate}

\subsection{Preference to Restrict the Post's Visibility}

One question read, ``Consider for yourself making a post with such photos in reality. With that in mind, please, continue the following statement with the option that would best describe you: 
 
\textbf{``Regarding the privacy settings of this post, ...''}

The response options were presented in randomized order:
\begin{itemize}[noitemsep]
    \item ...I will post it without restrictions.
    \item ...I will post it visible only for the users of the social network.
    \item ...I will post it visible only for friends and friends of friends.
    \item ...I will post it visible only for friends.
    \item ...I will post it visible only to certain groups of friends.
    \item ...I will post it visible only to me.
    \item ...I won't post it on my social network.
\end{itemize}

\subsection{Preference to Confirm Posting with Involved Friends}

One question read, ``Consider again for yourself making a post with such photos in reality. With that in mind, please, continue the following statement with the option that would best describe you: 
 
\textbf{``Regarding my friends' approval for this post, ...''}

The response options were presented in randomized order:
\begin{itemize}[noitemsep]
    \item ...I will post without asking anyone for approval.
    \item ...I will only post it, if certain friends present in the photos agree.
    \item ...I will only post it, if most of my friends present in the photos agree.
    \item ...I will only post it, if all my friends present in the photos agree.
    \item ...I won't post it, even if my friends present in the photos agree.
\end{itemize}

\subsection{Affective State (Experiment 1)}

One open-ended question read, ``Thinking about the task you were asked to imagine performing, \textbf{could you tell us briefly how you felt when you saw the PromotMe notice and what your feelings about publishing the post were?} If you want, instead of full sentences you can use descriptive adjectives.''

\subsection{Affective State (Experiment 2)}
\label{sec:app:DVs:affect:2}

Twenty items adapted from \citet{watson1988PANAS} to measure the valences of the positive affect (10 items) and negative affect (10 items).

\subsubsection*{Participant instructions:} When the PANAS is shown for the first time (before the experiment), ``You will now see 20 words (5 words per page) that describe different feelings and emotions. Please, read each item and \textbf{indicate to what extent you feel this way right now, at the present moment}.''

When the PANAS is shown for the second time (after the experiment, but before taking measures of the covariates), ``You will now \textbf{again} see \textbf{the same 20 words} (5 words per page) that describe different feelings and emotions. Please, read each item and \textbf{indicate to what extent you feel this way right now, at the present moment}.''

Both times on each page containing 5 items, the reminder read, ``\textbf{To what extent do you feel each of the emotions and feelings listed below, at the present moment?}''

\subsubsection*{Item statements:}
\begin{enumerate}[noitemsep]
    \item Interested.
    \item Distressed.
    \item Excited.
    \item Upset.
    \item Strong.
    \item Guilty.
    \item Scared.
    \item Hostile.
    \item Enthusiastic.
    \item Proud.
    \item Irritable.
    \item Alert.
    \item Ashamed.
    \item Inspired.
    \item Nervous.
    \item Determined.
    \item Attentive.
    \item Jittery.
    \item Active.
    \item Afraid.
\end{enumerate}

Items (1), (3), (5), (9), (10), (12), (14), (16), (17), (19) measure positive affect. Items (2), (4), (6), (7), (8), (11), (13), (15), (18), (20) measure negative affect.

\subsubsection*{Rating scale and anchoring.}
Five-point Likert-type rating scale anchored: 

\noindent Very slightly or not at all -- A little -- Moderately -- Quite a bit -- Extremely.

\section{Scales Measuring Individual Differences}
\label{sec:app:ind_diff}

\subsection{Curiosity and Exploration Inventory-II}
\label{sec:app:ind_diff:CEIII}

Ten items adapted from \citet{Kashdan2009curiosity} to measure two-dimensional curiosity.

\subsubsection*{Participant instructions:} ``On the next pages, you'll see two sets of statements with 10 statements in each set.

Please rate for \textbf{how accurately} the following statements reflect the way you generally feel and behave. Do not rate what you think you should do, or wish you would do, or things you no longer do. Please be as honest as possible.''

\subsubsection*{Item statements:}
\begin{enumerate}[noitemsep]
    \item I actively seek as much information as I can in new situations.
    \item I am the type of person who really enjoys the uncertainty of everyday life.
    \item I am at my best when doing something that is complex or challenging.
    \item Everywhere I go, I am out looking for new things or experiences.
    \item I view challenging situations as an opportunity to grow and learn.
    \item I like to do things that are a little frightening.
    \item  I am always looking for experiences that challenge how I think about myself and the world.
    \item I prefer jobs that are excitingly unpredictable.
    \item I frequently seek out opportunities to challenge myself and grow as a person.
    \item I am the kind of person who embraces unfamiliar people, events, and places.
\end{enumerate}

Items (1), (3), (5), (7), (9) measure \textit{stretching} dimension. Items (2), (4), (6), (8), (10) measure \textit{embracing} dimension. 

\subsubsection*{Rating scale and anchoring.}
Five-point Likert-type rating scale anchored: 

\noindent Very slightly or not at all accurate -- A little accurate -- Moderately accurate -- Quite a bit accurate -- Extremely accurate.

\subsection{Rational-Experiential Inventory-10}
\label{sec:app:ind_diff:REI10}

Ten items adapted from \citet{Epstein1996REI10} to measure the Need for Cognition (rational cognitive style, 5 items) and Faith in Intuition (experiential cognitive style, 5 items).

\subsubsection*{Participant instructions:}
``Next, you'll see the second set of 10 statements.

Rate these statements for \textbf{how truthfully or falsely} they reflect the way you generally feel and behave. Do not rate what you think you should do, or wish you would do, or things you no longer do. Please be as honest as possible.''

\subsubsection*{Item statements:}
\begin{enumerate}[noitemsep]
    \item I don't like to have to do a lot of thinking. [Reversed item]
    \item I try to avoid situations that require thinking in depth about something. [Reversed item]
    \item I prefer to do something that challenges my thinking abilities rather than something that requires little thought.
    \item I prefer complex to simple problems.
    \item Thinking hard and for a long time about something gives me little satisfaction. [Item excluded after reliability analysis]
    \item I trust my initial feelings about people.
    \item I believe in trusting my hunches.
    \item My initial impressions of people are almost always right.
    \item When it comes to trusting people, I can usually rely on my "gut feeling".
    \item I can usually feel when a person is right or wrong even if I can't explain how I know.
\end{enumerate}

Items (1) through (5) measure the rational cognitive style. Items (6) through (10) measure the experiential cognitive style.

\subsubsection*{Rating scale and anchoring}
Five-point Likert-type rating scale anchored:

\noindent Completely false -- Somewhat false -- Not sure -- Somewhat true -- Completely true

\subsection{Rational-Experiential Inventory-40}
\label{sec:app:ind_diff:REI40}

Twenty items adapted from \citet{pacini1999REI40} to measure the \textit{Rational ability} (rational cognitive style, 10 items) and \textit{Experiential ability} (experiential cognitive style, 10 items).

\subsubsection*{Participant instructions:}
``Next, you'll see the second set containing 20 statements.

Rate these statements for \textbf{how truthfully or falsely} they reflect the way you generally feel of yourself and behave. Do not rate what you think you should do, or wish you would do, or things you no longer do. Please be as honest as possible.''

\subsubsection*{Rational ability item statements:}
\begin{enumerate}[noitemsep]
    \item I'm not that good at figuring out complicated problems. [Reversed item]
    \item I am not very good at solving problems that require careful logical analysis. [Reversed item]
    \item I am not a very analytical thinker. [Reversed item]
    \item Reasoning things out carefully is not one of my strong points. [Reversed item]
    \item I don't reason well under pressure. [Reversed item]
    \item I am much better at figuring things out logically than most people.
    \item I have a logical mind.
    \item I have no problem thinking things through carefully.
    \item Using logic usually works well for me in figuring out problems in my life.
    \item I usually have clear, explainable reasons for my decisions.
\end{enumerate}

\subsubsection*{Experiential ability item statements:}
\begin{enumerate}[noitemsep]
    \item I don't have a very good sense of intuition. [Reversed item]
    \item Using my gut feelings usually works well for me in figuring out problems in my life.
    \item I believe in trusting my hunches.
    \item I trust my initial feelings about people.
    \item When it comes to trusting people, I can usually rely on my gut feelings.
    \item If I were to rely on my gut feelings, I would often make mistakes. [Reversed item]
    \item I hardly ever go wrong when I listen to my deepest gut feelings to find an answer.
    \item My snap judgments are probably not as good as most people's. [Reversed item]
    \item I can usually feel when a person is right or wrong, even if I can't explain how I know.
    \item I suspect my hunches are inaccurate as often as they are accurate. [Reversed item]
\end{enumerate}

\subsubsection*{Rating scale and anchoring}
Five-point Likert-type rating scale anchored:

\noindent Completely false -- Somewhat false -- Not sure -- Somewhat true -- Completely true

\section{Questionnaires of Experiences}
\label{sec:app:experiences}

\subsection{Privacy Violations}

The question read, ``Have you ever experienced any of the following? Please, select all that apply to you.'' 

The response options were presented in randomized order:
\begin{itemize}[noitemsep]
    \item Identity theft.
    \item Credit card fraud online.
    \item Online bullying.
    \item Online stalking.
    \item Any form of discrimination based on profiling, behavioral analytics, credit scoring, etc.
    \item Personal data breach (by an adversary attacking a company/person/government, whom you entrusted your data).
    \item Personal information released or sold online without your consent (by a company/person/government, whom you entrusted your information).
    \item Election fraud or voter rights violations based on information obtained or processed by computer systems.
    \item Personal data leak caused by a glitch in a computer system.
    \item Manipulating your choices or opinions through online profiling, behavioral analytics, etc.
    \item Tracking your location/activities by a state or federal government without your consent via the use of online computer systems (unless you were a suspect in a criminal investigation and these actions were authorized by the court at the time).
    \item Tracking your location/activities by a company without your consent via the use of online computer systems.
    \item Having your personal data lost by a company/person/government, whom you entrusted it.
    \item Any other misuse of your personal data, or any other abuse based on your personal data: [open-ended option].
\end{itemize}

\subsection{Types of Warnings and Notifications}

The question read, ``Which types of pop up indications do you recall encountering in the last month? (Select all that you can recall)''. 

The response options were presented in randomized order (except the last two):
\begin{itemize}[noitemsep]
    \item ``This website is using cookies''.
    \item ``This webpage is unsafe / its SSL certificate outdated''.
    \item ``This website wants to know your location''.
    \item ``This website wants to send you notifications''.
    \item ``This website tries to extract HTML5 canvas image data''.
    \item ``This page tried to open a pop-up window / Pop-up blocked''.
    \item ``We've updated our privacy policy / Read our privacy policy''.
    \item ``Are you sure you want to quit without saving? All progress may be lost''.
    \item ``This app requests access to...''
    \item ``Do you want to save password?''
    \item ``This connection is not private''
    \item ``Your session has expired''.
    \item ``Your login/password is incorrect''.
    \item Indications from antivirus and/or firewall software.
    \item Indications about system/software updates to be downloaded and/or installed.
    \item Other (you can specify one or more): [open-ended option].
    \item I don't remember encountering any in the last month.
\end{itemize}

\newpage 
\section{Suggested Posts}
\label{sec:app:suggested_post}

\begin{figure}[ht!]
     \centering
     \subfloat[Web view]{\includegraphics[width=.76\textwidth]{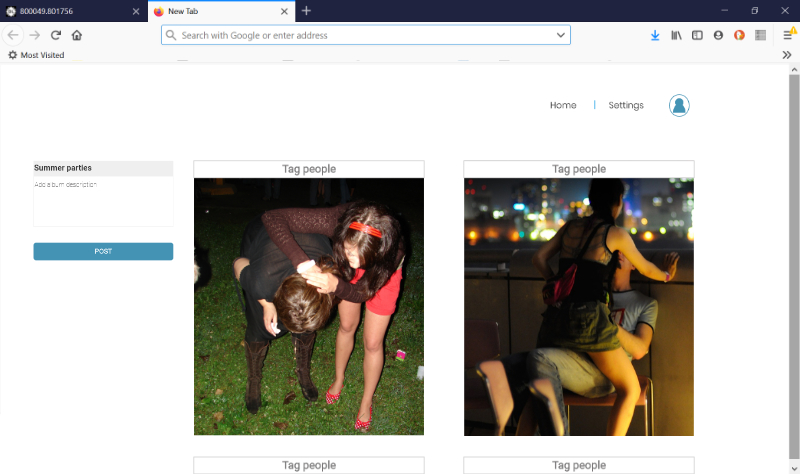}}\hspace{10pt}
     \subfloat[Mobile device view]{\includegraphics[width=.17\textwidth]{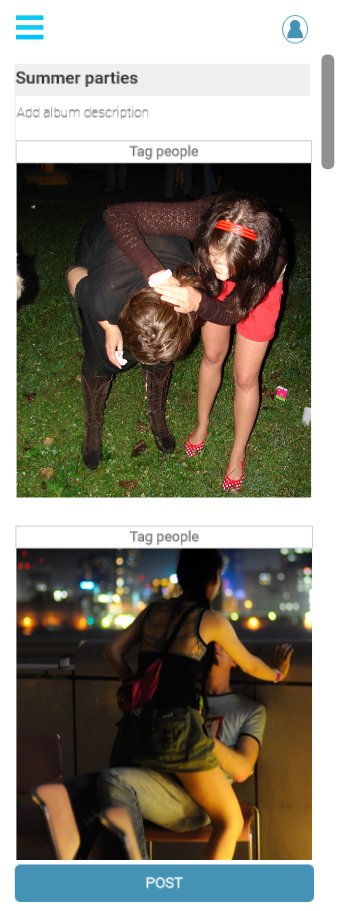}}
     \caption{The potentially sensitive suggested post, used in both experiments, as shown on different devices.}
     \label{fig:suggested_post}
\end{figure}

\begin{figure}[hb!]
     \centering
     \subfloat[Web view]{\includegraphics[width=.76\textwidth]{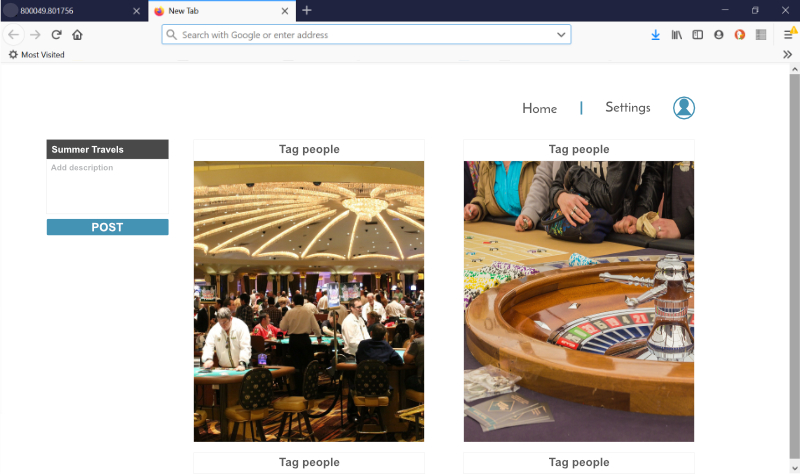}}\hspace{10pt}
     \subfloat[Mobile device view]{\includegraphics[width=.17\textwidth]{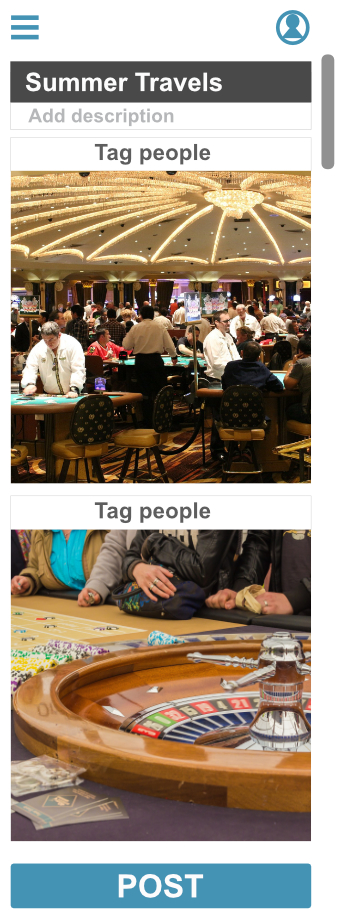}}
     \caption{The additional suggested post with potentially sensitive information, used in the second experiment, as shown on different devices.}
    \label{fig:suggested_post:gambling}
\end{figure}

\pagebreak

\begin{figure}[ht!]
     \centering
     \subfloat[Web view]{\includegraphics[width=.76\textwidth]{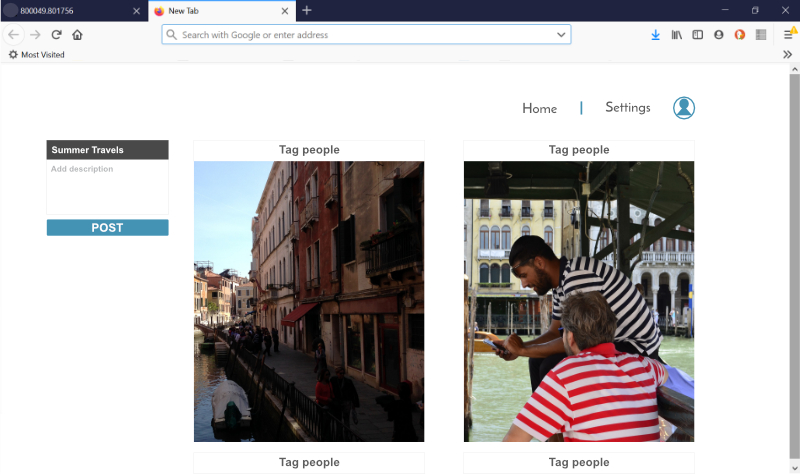}}\hspace{10pt}
     \subfloat[Mobile device view]{\includegraphics[width=.17\textwidth]{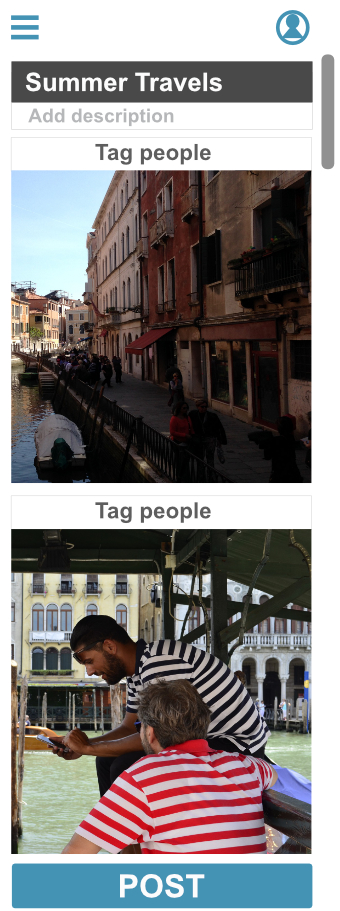}}
    \caption{The additional suggested post with neutral information, used in the second experiment, as shown on different devices.}
    \label{fig:suggested_post:traveling}
\end{figure}

\section{Experimental Scenario}
\label{sec:app:exp_scenario}

\subsection{Scenario Instructions}

First, the instructions describing the scenario were the same in both experiments. The instructions read,

``Imagine you have an account on a popular online social network. You also \textbf{have an app called PromotMe that helps you with posting} on the social network by automatically suggesting posts based on your recent activities.

Imagine also that \textbf{you've just returned from a trip} abroad, visiting friends and partying with them.
 
You have family members, colleagues, and friends connected.
 
Please, answer the questions that follow.''

\subsection{Suggested Post Instructions}

Next, any of the suggested posts (Appendix~\ref{sec:app:suggested_post}) when shown were accompanied with the following text: 

``The \textbf{PromotMe app suggests} setting up an album to \textbf{upload your photos} from the trip \textbf{on your social network} account:''

\end{document}